\newcommand{\pa}{\partial}
\newcommand{\be}{\begin{equation}}
\newcommand{\ee}{\end{equation}}
\newcommand{\bea}{\begin{eqnarray}}
\newcommand{\eea}{\end{eqnarray}}
\def \ci{\cite}

\newcommand{\bt}[1]{{\bar t}}

\documentclass[12pt]{article}

\usepackage{epsfig}
\input epsf

\def \sql {{\sqrt{\l}}\ }
\def \del{\partial}
\def \a {\alpha}
\def \aa {{\a'}}
\def\g{\gamma}
\def\s{\sigma}
\def\z{\zeta}
\def\zi{\zeta_1}
\def\zii{\zeta_2}
\def\ov{\over}
\def\la{\label}
\def\I{{\cal I}}
\def\J{{\cal J}}

\def\LL{{\cal J }}
\def \jL {{\rm J}}

\def\E{{\cal E}}
\def\w{\omega}
\def\b{\beta}
\def\l{\lambda}
\def\eps{\epsilon}

\def \adss{$AdS_5 \times S^5$}

\def \r { \rho}
\def \sql {\sqrt{\lambda} }

\def \vp {\varphi}

\def \ov {\over}
\def \s{\sigma}
\def \pa{\partial}

\def \ha {{1 \over 2}}

\def \la{\label}
\def  \Jt {  {J}_{\rm tot}    }

\def \k {\kappa}
\def\foot{\footnote}
\def \di{ \Xi}
\def \ff {\ {\rm f}}
\def \const {{\rm const}}

\def\K{\mbox{K}}    
\def\eE{\mbox{E}}   

\newcommand{\rf}[1]{(\ref{#1})}

\renewcommand{\thefootnote}{\fnsymbol{footnote}}

\def\appendix#1{
  \addtocounter{section}{1}
  \setcounter{equation}{0}
  \renewcommand{\thesection}{\Alph{section}}
  \section*{Appendix \thesection\protect\indent \parbox[t]{11.15cm}
  {#1} }
  \addcontentsline{toc}{section}{Appendix \thesection\ \ \ #1}
  }

\newcommand{\eq}[1]{(\ref{#1})}

\def\det{\hbox{det}}
\def\be{\begin{equation}}
\def\ee{\end{equation}}

\def \ci {\cite}

\def \foot {\footnote}
\def \bi{\bibitem}

\def \ha {{1 \over 2}}
\def \td {\tilde}

\def \ci{\cite}
\def \N {{\cal N}}
\def \ww {\Omega}

\begin{document}


\null\vskip-24pt \hfill AEI-2003-059
\vskip-1pt
\hfill
Imperial/TP/2-03/28
\vskip-1pt
\hfill {\tt hep-th/0307191}
\vskip0.2truecm
\begin{center}
\vskip 0.2truecm {\Large\bf
Spinning strings in $AdS_5\times S^5$\\
\vskip 0.2truecm
and integrable systems
}
\\
\vskip 0.5truecm
{\bf G. Arutyunov$^{1,\star,}$\foot{emails: {
agleb@aei-potsdam.mpg.de;\  jrusso@ecm.ub.es;\
frolov, tseytlin @mps.ohio-state.edu
}
\newline
$^{\star}$On leave of absence from Steklov Mathematical
 Institute,
Gubkin str.8,
117966, Moscow, Russia},
S. Frolov$^{2,\star}$,
J. Russo$^{3,4}$
and A.A. Tseytlin$^{2,3,}$\footnote{Also at Lebedev
 Institute, Moscow.}\\
\vskip 0.4truecm
$^{1}$
{\it Max-Planck-Institut f\"ur Gravitationsphysik,
Albert-Einstein-Institut, \\
Am M\"uhlenberg 1, D-14476 Golm, Germany}\\
\vskip .2truecm
$^{2}$ {\it Department of Physics,
The Ohio State University,\\
Columbus, OH 43210-1106, USA}\\
\vskip .2truecm
$^{3}$
{\it
Blackett Laboratory, Imperial College, London, SW7 2BZ, U.K.}\\
\vskip .2truecm
$^{4}$ {\it Departament ECM,
Facultat de F\'\i sica, Universitat de Barcelona\\
Instituci\' o Catalana de Recerca i Estudis Avan\c{c}ats (ICREA),
 Spain} }
\end{center}
\vskip 0.2truecm \noindent\centerline{\bf Abstract}
\vskip .2truecm
We show that solitonic solutions
 of the classical string
action on the \mbox{$AdS_5\times S^5$} background
that carry
charges (spins) of the Cartan subalgebra of the
 global symmetry   group
can be classified in terms of  periodic solutions of
 the Neumann
integrable system. We derive equations which determine the
energy of these solitons  as a function of spins.
In the limit
of  large spins $J$, the first subleading $1/J$ coefficient  in
the  expansion of the string energy is expected  to be  non-renormalised
to all orders in the inverse string tension
expansion and thus can be directly
compared to the 1-loop anomalous dimensions of the corresponding
composite operators in ${\cal N}=4$ super YM theory.
We obtain a closed system of equations
that  determines this subleading
coefficient  and, therefore, the 1-loop
anomalous dimensions of the dual SYM operators.
We  expect that an equivalent  system of
equations should follow from the thermodynamic limit of the
algebraic Bethe ansatz for
the  SO(6) spin chain derived
from  SYM theory.
{}We also identify a particular string solution whose classical
energy exactly reproduces
the one-loop anomalous dimension of a certain set of SYM operators
with two independent R charges $J_1, J_2$.


\newpage

\renewcommand{\thefootnote}{\arabic{footnote}}
\setcounter{footnote}{0}

\section{Introduction and summary}

Recently,  there was a remarkable
progress towards  understanding
AdS/CFT  duality in non-supersymmetric sector of states
of string theory on \adss\
\ci{ft2,ft3,mz2,ft4},
 generalizing  earlier  work of
\ci{bmn,gkp,ft1}.
On the string side, one
identifies semiclassical states
described by  solitonic  closed-string solutions
on a 2-cylinder. They have
 finite energy and   carry
  SO(6) (and,  in general, SO(2,4))
angular momentum  components $J_i$.
 In the limit of large
angular momenta  the first subleading
term in the  expansion of the classical
string energies
happens to be protected  (i.e. is not
renormalised by string $\a'$ corrections)
and thus can be matched onto the dimensions
of the corresponding gauge-theory ($\N=4$ SYM)
operators \ci{mz2}.
We refer to \ci{ft4} for a more detailed discussion.

In general,
 according to the AdS/CFT duality the \adss\,  string sigma
model
should be
equivalent to the ${\cal N}=4$ supersymmetric SU($N$) YM theory
with $N\to \infty$.
The  composite primary operators in this theory are classified
in
terms of
UIR's of the superconformal group PSU(2,2,$|$4), i.e. by the
conformal dimension $\Delta$,  two spins $S_1,S_2$
and by the Young tableaux
(or,
equivalently,  by the Dynkin labels) of the R-symmetry
group SU(4).
At $N=\infty$ only single-trace operators matter. Thus
one should expect that
the energy of the string solutions
 considered as the function
of the angular momenta
 $J_i$  should match with the dimensions of the corresponding
primary single-trace operators in the  SYM  theory.

The bosonic part of the  classical string action is
a combination of SO(2,4) and SO(6)   sigma models.
The O(n) (or O(p,q))  sigma  models are
known to be  classically integrable \ci{lup},
and the same should obviously be  true also
upon imposition of the conformal gauge constraints, i.e.
for the corresponding classical string theories
(for some  related work see \ci{BN,veg,man}).
One expects, therefore,
a close connection  between special classes of  string solutions
representing particular semiclassical string states  and
certain   integrable models.
 As was already observed  earlier,
  the folded rotating string  solutions with
one \ci{vog,gkp} or two \ci{ft1,ft4} non-vanishing
angular momenta  are related to the
1-d sine-Gordon model.

Here we will consider a
generalization to the case  when
all  the three ``Cartan''
components of the  SO(6) angular momentum
 are  non-zero and will
find that in this case
the  SO(6)  sigma model effectively reduces to a special
 integrable 1-d model --  the
{\it Neumann model} \ci{N}. The latter    describes  a three-dimensional
harmonic oscillator with three different frequencies
constrained to move
on a two- sphere
(see, e.g., \ci{per,bab,mam}).

The class of $S^5$
rotating string solutions we  will be discussing is
 parametrized
by the  angular momenta $J_i=(J_1,J_2,J_3)$
with the energy being $E=E(J_1,J_2,J_3)$.
To be able to compare to  perturbative conformal dimensions on the
 SYM side one  needs to assume that
\be \frac{\lambda}{  J_i^2}  \ll 1 \ , \ \ \ \ \ \ \ \
\frac{1}{ J_i }\ll 1 \ ,  \la{asu}
\ee
where $\l$ is the square of string tension (related
as usual to `t Hooft
coupling).
In this case  the \adss\ superstring
$\a'$ corrections to the classical energy
will be suppressed by extra powers of $1\ov J_i$.
This happens \ci{ft3}
despite the non-BPS nature  of the extended
rotating string states (the only BPS state is a point-like string
having only one non-zero component of $J_i$ \ci{bmn})
and is due to the fact that the underlying
superstring theory has (i) global supersymmetry and (ii)
is effectively massive in this case (with 2-d masses $\sim {1\ov J_i}$)\foot{
A similar argument
explains \ci{ft1,semi}
why 2- and higher loop corrections
to the energy of string states in the BMN \ci{bmn} sector are suppressed
by powers of $1/J$.}.

Assuming \eq{asu},
the  classical energy can  be expanded as\foot{In the
cases when there is  just one non-vanishing component of the spin
as in \ci{gkp} or the conditions \eq{asu}
are not satisfied for at least one of the spins \ci{ft1,jr},
the energy may contain  also a constant subleading term $O(\sql)$.
Under the condition $ {\sql \ov J} \ll 1 $ such   term
(which will not be protected against string
 $\a'\sim { 1 \ov \sqrt{\l}}$-corrections
 and thus cannot be easily compared to SYM theory)
is much larger that the subleading term in the equation below,
and thus the corresponding SYM  operators should have larger
anomalous dimensions.}
\be\la{eee}
E= \Jt + {\lambda \ov \Jt}
f_1({J_i\ov \Jt}) + {\lambda^2 \ov \Jt^3}
f_2({J_i\ov \Jt})+ ... \ ,
\ \ \ \ \
\ee
\be   \Jt \equiv  J_1 + J_2 + J_3 \ . \ee
One  immediate  aim is then
to determine   the coefficient function
$f_1({J_2\ov \Jt}, {J_3\ov \Jt})$.
Given the analytic dependence of the subleading term in $E$
on $\l$ and its expected ``non-renormalizability'' on the string side,
one may be able, as explained
 in \ci{ft3,ft4},   to compare it directly
to the one-loop anomalous dimensions
of gauge-theory operators  of the type
tr$[ (\Phi_1 + i \Phi_2) ^{J_1}  (\Phi_3 + i \Phi_4) ^{J_2}
(\Phi_5 + i \Phi_6) ^{J_3}]$+...
belonging to irreducible representation
of SU(4)   with Dynkin labels
$[J_2-J_3, J_1-J_2, J_2 + J_3]$
(we assume for definiteness that $ J_3 \leq J_2 \leq
J_1$).\foot{The primary operators obtained after diagonalizing
the dilatation matrix for the gauge-invariant operators
mentioned above
are not {\it super}conformal primaries lying on the unitary bound
of the continuous (unprotected)
series of UIR's of PSU(2,2$|$4), as can be seen from the
corresponding relation
between the conformal
dimension and the Dynkin labels. Their parent superconformal
primaries can be found by analysing representations of supersymmetry.
However, this is immaterial since the anomalous dimensions of
 primaries
and of their susy descendants are equal.}
Note also that the expected non-renormalisation of the $1/\Jt$ term
to all orders in the inverse string tension predicts (through
AdS/CFT duality)
that the corresponding term in ${\cal N}=4$ SYM  should be one loop exact.

Finding the spectrum of one-loop  anomalous dimensions
of such scalar operators with all three  $J_i$
being non-zero
should be possible,
as  in the two-spin case ($J_3=0$)
in \ci{mz2}, using
the techniques
(dilatation operator related to integrable
 spin chains  and Bethe ansatz)
developed in a recent remarkable series of papers
\cite{mz1,bes1,mz2,bes2}.
Here we will determine $f_1$ in several special cases
with $J_i\not=0$,
thus making
string-theory  predictions  for
the corresponding    eigenvalues of the anomalous dimension
 matrix.

One special
 three-spin solution was found  already  in \ci{ft2,ft3}:
this  is a circular string  with
 $J_1=J_2$  and arbitrary $J_3$
(the stability condition implies
$J_1 + J_2  \leq { 4 n -1 \ov ( 2n-1)^2 } J_3$ where $n$ is
the winding number; in what follows  we  set $n=1$)
$$
E= J_1 + J_2 + J_3 + {\lambda   (J_1 + J_2)  \ov 2
(   J_1 + J_2 + J_3  )^2 } + ... $$
\be\la{bah} =\
\Jt + {\lambda     \ov
2 \Jt  } \  ({ 1 } - { J_3 \ov \Jt } )
+ ...
 \ ,
\ee
where dots stand for $\l^2$ and other subleading  terms.
The special case  of $J_3=0$, i.e.
\be\la{hah}
E= J_1+ J_2 + f_1 {\lambda      \ov
J_1+ J_2 } + ... \  ,  \ \ \ \ \  \ \ \ \ \ \ \
 f_1 = { 1 \ov 2}
\ee
 corresponds to  the  simplest
 two-spin {\it circular}
  string   solution \ci{ft2}.
 In spite of being unstable, this solution has its ``counterpart''
 on the  SYM side  \ci{mz2}.
 Another   string state   with the same
 quantum numbers $J_1=J_2,\ J_3=0$
 but lower energy is represented  by the stable
 {\it} folded
 string solution \ci{ft4} (which generalizes the
 single-spin solution of
 \ci{gkp} to the two-spin  case).
 In this case the energy is
 given
 by \rf{hah} with\foot{This number has a simple origin
in terms of values of elliptic functions as mentioned at the end of Section 3.3.}
 $$f_1= 0.356... \ ,$$
  and, remarkably,
 can be matched  exactly
  with the corresponding lowest anomalous dimension eigenvalue
 on the SYM side  \ci{mz2}.

 As in the two-spin case, in  general, there will
 be several three-spin string solutions
 for given values  of $J_1,J_2,J_3$
 having different values of $E$. The first subleading term in $E$
 will then  be expected to correspond to
 the band of one-loop dimensions of  the
 SYM eigen-operators in the $[J_2-J_3, J_1-J_2, J_2 + J_3]$
  irrep.
 In particular,  there may be several string solutions with
 $J_1=J_2$ and small $J_3$  generalizing the   circular \ci{ft2,ft3}
 and folded \ci{ft4}
 two-spin solutions, and different from the circular string
 solution of \ci{ft3} with  $E$ given in \rf{bah}.

 We shall see that in spite of the formal integrability
 of the Neumann model,
 finding the explicit form of the three-spin  solutions and,
  in particular, their
 energies, turns out to be  complicated.
 Below  we shall concentrate  on
 several special cases.
 In particular,  there are two obvious
 cases  generalizing the two-spin solutions mentioned above:
 (i) generalization of the folded two-spin solution
 to the case of non-zero $J_3  < J_1=J_2$;
 (ii) generalization of the circular  two-spin solution
 to the case of non-zero $J_3  < J_1=J_2$
 which has less energy than the circular three-spin solution
 of \ci{ft3} (the latter
  is unstable  for small $J_3$).
 In these and similar  cases with $J_1,J_2 \gg J_3$
   we find the    following expression for the
 energy (to the leading order in $J_3\ov \Jt$)
 \be \la{eqq}
E= \Jt + {\lambda \ov \Jt}
( f^{(0)}_1   +  f^{(1)}_1 { J_3 \ov \Jt } + ...)  + ...  \ .
\ \ \ \ \
\ee
 Note that the  expression \eq{bah}
for  the circular three-spin solution of \ci{ft3} is thus a
 special case of \rf{eqq}.\foot{
 Let us mention that a
 reason for  considering linear in $J_3$ terms in $E$ (i.e.
 leading
deformations of  the   two-spin expressions)
 is that the
corresponding leading terms in SYM anomalous
 dimensions may be possible
 to compute by using a perturbation
 theory near the Heisenberg model Hamiltonian
 corresponding to the two-spin case,
 i.e. to the  anomalous dimension matrix
 for the operators in the $[J_2,J_1-J_2,J_2]$ irrep.}
 One of our  aims will be to compute the value of the
coefficient $ f^{(1)}_1 $ for various
three-spin solutions.
In particular, we will find  in Section 3 that
the folded string solution that generalizes  the
$J_1=J_2,\ J_3=0$ solution of \ci{ft4}  has
$f^{(1)}_1= 4.79...$.

One may try  to find also folded string solutions with
$J_1=J_2=J_3$, which should have less energy than the circular
solution of \ci{ft3}. Though the  latter  is stable for
$J_1=J_2=J_3$, it is likely to be only a local minimum
of the energy, i.e. there   may be  another
$J_1=J_2=J_3$  solution with less energy.\foot{
In general, there may be several local minima,
i.e. stable solutions with the same quantum numbers.}

The rest of the paper is organized as follows.
In Section 2 we shall present the ansatz for the general three-spin
$S^5$  rotating string solution and explain its relation to
the Neumann integrable system. This  will allow us to reduce the
problem to a pair of first-order differential equations
for the two coordinates of $S^2$ related to 5-th order polynomial
defining a hyperelliptic curve of genus 2.

In Section 3.1 we shall argue that to obtain
a non-trivial folded string solution with
the three non-zero spins the string must be ``bent''
(i.e. two coordinates of $S^2$ should have a different
number of folding points).
 In Section 3.2 we shall  derive the
general system  of equations that governs the form
of the subleading (or ``one-loop'') term  $f_1$ in the expression
\rf{eee} for the energy of the bent string.
An  equivalent system  is expected to follow
 from the thermodynamic limit of the algebraic
Bethe ansatz for the SO(6) spin chain
determining the one-loop anomalous dimensions on the SYM side
\ci{mz1,bes2}.
In Section 3.3  we shall study this system in expansion in small $J_3$
and determine the coefficient $f_1^{(1)}$ in \rf{eqq}
in the special case of perturbation near  two-spin folded string solution
of \ci{ft4} with  $J_1=J_2$.


Section 4 will be devoted to a different class of three-spin solutions
which  will have higher energy than folded bent strings
 for the same
values of $J_i$.
Using a combination of analytic and numerical methods
we shall again  determine the form of the
leading correction $f_1$ in \eq{eee} in this case.

In Section 5 we shall
consider a  two-spin solution of a circular type that generalizes
the circular  solution of \ci{ft2} to the case of unequal $S^5$ spins
$(J_1,J_2)$. We shall show that, just in the case of the two-spin folded
string in \ci{ft4}, the
first subleading term in the corresponding expression for the energy
matches precisely the one-loop anomalous dimensions of a
set of  SYM operators with SU(4) Dynkin labels $[J_2,J_1-J_2,J_2]$
which
correspond to solutions of the Bethe ansatz equations in \ci{mz2}
 with all Bethe roots
lying on the imaginary axis.
This complements the results  in \ci{mz2,ft4},  providing another   remarkable
 test of the AdS/CFT correspondence.

Similar solutions describing
string spinning in $AdS_5$ directions
can be analysed  in much the same way as described in Section 6.
In fact, most of the  SO(6) case equations have a
direct analog  in the SO(2,4) case
or are related  by an analytic continuation.

In Appendix A we shall explain how the general solution of the Neumann
model can be written in terms of $\theta$-functions defined on the Jacobian
of a hyperelliptic genus 2 Riemann  surface \ci{mam}.
Appendix B  will contain a list of integrals used in Section 3.
In Appendix C we shall study the vanishing of
other (``non-Cartan'') components of the SO(6) angular momentum
tensor for different three-spin solutions  which is crucial
\ci{ft2} for their consistent semiclassical
quantum state interpretation
(and thus a possibility to  establish a correspondence with
particular SYM operators with the same quantum numbers).
In Appendix D
we shall describe the two-spin solution corresponding to the straight  folded
string without bend points. We will show that
such string solution does not allow a deformation towards a non-zero
 third spin component.

\setcounter{equation}{0}

\section{Reduction of  O(6) sigma-model to the \\ Neumann system}

\subsection{Rotating string ansatz and integrals of motion }

Let us
consider the bosonic part of the classical closed
string propagating in the
$AdS_5\times S^5$ space-time. The world-sheet action
in the  conformal gauge is
\be
I= - { \sql  \ov 4\pi }
\int d\tau d\sigma  \ \big[ G^{(AdS_5)}_{mn}(x)
\del_a x^m \del^a  x^n\ + \    G^{(S^5)}_{pq}(y)  \del_a y^p
\del^a y^q \big] \ , \ \ \ \ \ \ \ \ \sql \equiv  { R^2 \ov
\aa} \ .  \la{A}
\ee
The two metrics have the standard form
in terms of the 5+5 ``angular''  coordinates:
\be \la{dam}
(ds^2)_{AdS_5}
= - \cosh^2 \r \ dt^2 +  d\r^2 + \sinh^2\r \ (d \theta^2 +
 \sin^2 \theta  \ d \phi^2 + \cos^2 \theta \ d\vp^2)  \ , \ee
 \be \la{adm}
 (ds^2)_{S^5}
= d\g^2 + \cos^2\g\ d\vp_3^2 +\sin^2\g\ (d\psi^2 +
\cos^2\psi\ d\vp_1^2+ \sin^2\psi\ d\vp_2^2)\ . \ee

It is convenient to represent \eq{A}
as an action for the O(6)$\times $SO(4,2)
sigma-model (we follow the notation of \ci{ft2})
\be
I={ \sql  \ov 2\pi }\int d\tau d\sigma (L_S+L_{AdS})\
, \la{Lsp} \ee
where
\bea
\la{SL}
L_S&=&-\frac{1}{2}\pa_a X_M\pa^a X_M+\frac{1}{2}\Lambda
(X_MX_M-1)\, , \\
L_{AdS}&=&-\frac{1}{2}\eta_{MN}\pa_a Y_M\pa^a Y_N+\frac{1}{2}
\td \Lambda (\eta_{MN}Y_MY_N+1)\, .
\la{SSL}
\eea
Here  $X_M$, $M=1,\ldots , 6$  and $Y_M$,
 $M=0,\ldots , 5$ are  the
the embedding coordinates of
${R}^6$
with the Euclidean metric in $L_S$ and
 with  $\eta_{MN}=(-1,+1,+1,+1,+1,-1)$
 in $L_{AdS}$ respectively.
 $\Lambda$  and $\td \Lambda$ are the Lagrange multipliers.
The action \eq{Lsp} is  to be supplemented
 with the usual conformal gauge  constraints.

The embedding coordinates are related to the
``angular'' ones in \eq{dam},\eq{adm} as follows:
\be\la{relx}
 X_1 + i X_2 = \sin   \g \ \cos
\psi \ e^{ i \vp_1} \ , \ \ \ \
X_3 + i X_4 =  \sin   \g \ \sin \psi \
e^{ i \vp_2} \ , \  \ \ \ \ \
X_5 + i X_6 = \cos  \g \ e^{ i \vp_3} \ ,
\ee
\be \la{rell}
Y_1+ iY_2 = \sinh \r \ \sin \theta \  e^{i \phi}\ , \ \ \ \ \
Y_3 + i Y_4  = \sinh \r \ \cos \theta \  e^{i  \vp}\  , \ \ \ \
Y_5 + i Y_0 = \cosh \r \  e^{i t } \ . \ee
In the next few sections we will
be discussing the case when the
string is located at the center of $AdS_5$
and  rotating in $S^5$, i.e.
is trivially embedded in $AdS_5$ as $Y_5+iY_0=e^{i\kappa \tau}$
with  $Y_1,...,Y_4=0$.

The $S^5$ metric
has three commuting translational
isometries in $\vp_i$ which give rise to three global
commuting integrals of motion (spins) $J_i$. Since we are
interested in the periodic motion with three $J_i$ non-zero
it is natural to choose the following ansatz for $X_M$:
\be
\la{emb}
X_1+iX_2=x_1(\sigma)\ e^{iw_1\tau}\, , ~~~
X_3+iX_4=x_2(\sigma)\ e^{iw_2\tau}\, , ~~~
X_5+iX_6=x_3(\sigma)\ e^{iw_3\tau}\, ,
\ee
where the real radial functions
 $x_i$   are independent of time
and  should, as a consequence of $X_M^2=1$,
 lie on a two-sphere $S^2$:
$$\sum^3_{i=1} x_i^2=1 \ , \ \ \ \ \ \ \  i=1,2,3 \ . $$
Then the  spins $J_1=J_{12}, \ J_2=J_{34}, \ J_3=J_{56}$
forming  a  Cartan subalgebra of  SO(6)  are
\be
\la{spins}
J_i=\sqrt{\lambda}\ w_i
\int_0^{2\pi}\frac{d\sigma}{2\pi}\
 x_i^2(\sigma)\equiv\sqrt{\lambda}\ {\cal J}_i \, .
\ee
As discussed  in \ci{ft2}, to have a consistent
semiclassical string state interpretation of these
configurations one should  look for solutions  for which
all other components of the SO(6) angular momentum tensor
$J_{MN}$ vanish.

The space-time energy $E$ of the string
 (related to a generator  of the
compact SO(2) subgroup of SO(4,2)) is simply
\bea \la{enn}
E=\sqrt{\lambda}\ \kappa\equiv \sqrt{\lambda}\ \E \, .
\eea
The only non-trivial
Virasoro constraint is then
(dot and prime are derivatives over $\tau$ and $\sigma$)
\bea
\kappa^2=\dot{X}_M\dot{X}_M+ X_M ' X_M '\, .
\eea
As a consequence of
 this relation the energy becomes a function of the SO(6) spins:
\bea
E=E(J_1,J_2,J_3)\, .
\eea
On the string theory side of the AdS/CFT duality
the problem is thus to  classify the  solutions of the
string sigma-model subject to the Virasoro constraints
with further determination of their space-time
energy as a function of the spins.
Below we shall derive a closed system of equations
which,  in principle,
allows one  to find the energy as a function of the spins
for generic three-spin solutions
and therefore to determine
the dimensions of the corresponding
gauge theory operators.

Substituting the ansatz  (\ref{emb})
into the SO(6) Lagrangian   \eq{SL}
 we get the following 1-d (``mechanical'')
 system
\be
\label{L}
L=\frac{1}{2}\sum^3_{i=1} (x'^2_i-w_i^2 x_i^2)+\frac{1}{2}
\Lambda(\sum^3_{i=1} x_i^2-1) \, .
\ee
It  describes  an $n=3$ dimensional
harmonic oscillator constrained to remain on a
unit $n-1=2$ sphere. This is the
special  case of the $n$-dimensional
 {\it Neumann}  dynamical
system \cite{N} which is known to be integrable
\ci{per}.

The Virasoro constraint
implies  that the energy $H$ of the Neumann
system is given by
\bea\la{hhh}
H=\ha \sum^3_{i=1 } (x'^2_i + w_i^2 x_i^2)
=\ha \kappa^2 \, .
\eea
Solving the equation of motion for the Lagrange multiplier
$\Lambda$ we obtain
the following non-linear equations for $x_i$:
\be
\label{Eqofm}
x_i''=-w_i^2x_i-x_i\sum_{j=1}^3
 \Big(x'^2_j-w_j^2x_j^2\Big) \, .
\ee
The canonical momenta conjugate to $x_i$ are
$$\pi_i=x_i' \ , \ \ \ \ \ \ \ \  \  \sum^3_{i=1} \pi_i x_i=0\ . $$
One can think about this dynamical system
as being originally defined on the cotangent bundle
$T^{*}{R}^3$
. Imposing the constraints
reduces the phase space
to $T^{*}S^2$.
The Dirac bracket obtained from the canonical structure
$\{\pi_i,x_j\}=\delta_{ij}$ is
\bea
\{\pi_i,\pi_j \}_{D}=x_i\pi_j-x_j\pi_i\, , \ \ \
\{\pi_i,x_j \}_{D}=\delta_{ij}-x_ix_j\, ,\ \ \
\{x_i,x_j \}_{D}=0 \, . \la{ps}
\eea
One can check  that  (\ref{Eqofm})
follows from  the  Poisson
structure (\ref{ps}) and the Hamiltonian
\bea
H=\frac{1}{2} \sum^3_{i=1} (\pi_i^2+w_i^2x_i^2)
\eea
supplemented with the two constraints
$ \sum^3_{i=1} x_i^2=1, \  \sum^3_{i=1} \pi_i x_i=0$.

The crucial point allowing to solve this  model
is that the $n$-dimensional ($n=3$ in the present case)
 Neumann system has the following
$n$  integrals of motion \cite{U}:
\bea \la{inti}
F_i=x_i^2+\sum_{j\neq i}\frac{(x_i\pi_j-x_j\pi_i)^2}
{w_i^2-w_j^2} \,  , \ \ \ \ \ \ \ \ \ \
\sum^n_{i=1}  F_i=1 \ .
\eea
In the present case  $n=3$ and thus
only {\it two}
 of the three  integrals  of motion  are independent.
Moreover, these integrals are in involution
with respect to the Poisson bracket (\ref{ps}) and
the Hamiltonian is
\be H=\frac{1}{2}\sum^3_{i=1} w_i^2F_i \ . \ee
Thus, any two of these three  integrals
of motion are enough to integrate this
dynamical system since  the motion
 occurs on a surface of constant integrals.

In order to find the relevant closed string  solutions
 we need also to impose
the periodicity conditions  on  $x_i$:
\be\la{pep}  x_i (\s) = x_i (\s + 2 \pi) \ , \ee
i.e. we are  interested in  ``periodic'' version of the
Neumann model.

\subsection{First-order system for the  ellipsoidal coordinates}

It is convenient to describe the
phase space of this  model
in terms of
independent 2+2 canonical variables rather than
the 3+3  constrained variables  $x_i,\pi_i$.
One natural coordinate system  on a two-sphere
is the angular $(\g,\psi)$
one implied by \eq{relx} and \eq{emb}:
\be  x_1 = \sin   \g \ \cos \psi  \ , \ \ \ \
x_2=  \sin   \g \ \sin \psi  \ , \  \ \ \ \ \
x_3  = \cos  \g \ . \la{jio} \ee
 However,
if all the frequencies
$w_i$ are different and so   the Hamiltonian is not spherically
 symmetric, it appears advantageous to
 use the  so called ellipsoidal coordinates
\cite{bab}.
The ellipsoidal coordinates are introduced
as the two real roots
$\zeta_1$ and $\zeta_2$ of the following
quadratic equation
\be\la{ddd}
\frac{x_1^2}{\zeta-w_1^2}+\frac{x_2^2}
{\zeta-w_2^2}+\frac{x_3^2}{
\zeta-w_3^2}=0 \, .
\ee
Assuming $w_1<w_2<w_3$ we can define the range of $\zeta_a$
($a=1,2$)
as
\be
\la{r}
w_1^2\leq \zeta_1\leq w_2^2\leq \zeta_2 \leq w_3^2 \, .
\ee
With this  range  $\zeta_a$
cover  $\frac{1}{8}$-th  of the two-sphere corresponding to
$x_i\geq 0$. One can think of the whole sphere as  covering
the domain (\ref{r}) and branching along its boundary.
For $x_i\geq 0$ we have
\be\label{xizi}
x_1=\sqrt{\frac{(\zeta_1-w_1^2)(\zeta_2-w_1^2)}{w_{21}^2
w_{31}^2}}\ , \ \ \ \ \
x_2=\sqrt{\frac{(w_2^2-\zeta_1)(\zeta_2-w_2^2)}{w_{21}^2
w_{32}^2}}\, , \ee
\be
x_3=\sqrt{\frac{(w_3^2-\zeta_1)(w_3^2-\zeta_2)}{w_{31}^2
w_{32}^2}} \, ,
\ \ \ \ \ \ \ \ \ \ w_{ij}^2\equiv w_i^2-w_j^2\ .  \ee
One can check  that
$\sum^3_{i=1} x_i^2=1$, i.e. we indeed  get a parametrization
of a two-sphere.
Substituting now this parametrization for $x_i$ into eq. (\ref{L})
we get the following sigma-model Lagrangian
\bea
\label{Lag}
L=\frac{1}{2}g_{ab}(\zeta)\ \zeta'_a \zeta'_b-U(\zeta)  \, ,
\eea
where the non-zero components of the two-sphere metric are
\be
g_{11}=\frac{\zeta_2-\zeta_1}{4(\zeta_1-w_1^2)(w_2^2-\zeta_1)
(w_3^2-\zeta_1)}\ , \ \ \
g_{22}=\frac{\zeta_2-\zeta_1}{4(\zeta_2-w_1^2)(\zeta_2-w_2^2)
(w_3^2-\zeta_2)} \ ,
\ee
and the potential $U$ is very simple
\bea
\la{poten}
U=\frac{1}{2}(w_1^2+w_2^2+w_3^2-\zeta_1-\zeta_2) \, .
\eea
Note that in the domain (\ref{r}) the metric
$g_{ab}$ is non-negative.

Expressing the integrals of motion \eq{inti}
in terms of $\zeta_a$ one finds a system of two 1-st order
equations which can also be obtained
by solving directly the associated Hamiltonian-Jacobi
problem
\bea
\la{sep}
\left(\frac{d\zeta_1}{d\sigma}  \right)^2=
-4\frac{P(\zeta_1)}{(\zeta_2-\zeta_1)^2}\, , ~~~~~~
\left(\frac{d\zeta_2}{d\sigma}  \right)^2=
-4\frac{P(\zeta_2)}{(\zeta_2-\zeta_1)^2}\, .
\eea
Here the function $P(\zeta)$ is a 5-th order polynomial
\bea \la{poll}
P(\zeta)=(\zeta-w_1^2)(\zeta-w_2^2)(\zeta -w_3^2)(\zeta -b_1)
(\zeta -b_2)\ .
\eea
The parameters
 $b_1,b_2$ are the
 two constants of motion which can be expressed in
  terms
of  integrals $F_i$ in \eq{inti}  by solving the system of equations
\bea
\nonumber
b_1+b_2&=&(w_2^2+w_3^2)F_1+(w_1^2+w_3^2)F_2+(w_1^2+w_2^2)F_3\, , \\
b_1b_2&=&w_2^2w_3^2F_1+w_1^2w_3^2F_2+w_1^2w_2^2F_3\, .
\eea
In terms of variables $b_i$ the Hamiltonian \eq{hhh} reads as
\bea \la{enna}
H=\frac{1}{2}\Big(w_1^2+w_2^2+w_3^2-b_1-b_2\Big) = \ha \k^2
= \ha \E^2  \, .
\eea
 In what follows we shall assume that
\be \la {bb}  b_1 \leq  b_2  \ . \ee
In this case \eq{sep}  implies that
\bea
\la{zibi}
b_1\le \zi\le b_2\ , \ \ \ \ \ \ \ \ \  b_2\le \zii\ .
\eea
Let us note also that the  polynomial $P(\zeta)$ in \eq{poll}
 can be interpreted  as
defining  a hyperelliptic curve of genus 2
\bea\la{ggg}
s^2+P(\zeta)=0 \, ,
\eea
with $s$ and $\z$ being two complex coordinates.
Thus, we have found  that
the most general three-spin string solutions
 are naturally associated  with  hyperelliptic curves.

The system \eq{sep} allows one to achieve the full separation of the
variables:
dividing one equation in (\ref{sep}) by the other
 one can integrate, e.g.,
$\zeta_2$ in terms of $\zeta_1$ and then obtain a closed equation
for $\zeta_1$ as the function of $\sigma$.
In finding the solutions
we  need also to take into account the periodicity
conditions \eq{pep}   now
viewed as conditions on $\z_1,\z_2$.

The spins $J_i = \sql \ \J_i$ in
\eq{spins} expressed in terms of $\z_1,\z_2$
satisfy the
 following relations
\be
\la{spi}
{\J_1\ov w_1}+ {\J_2\ov w_2}+{\J_3\ov w_3}=1\ ,
\ee
\be
\la{spii}
w_1 \J_1 +w_2 \J_2 +w_3 \J_3 = w_1^2+w_2^2+w_3^2 -
\int_{0}^{2\pi}\frac{d\sigma}{2\pi} \ (\zeta_1+\zeta_2) \ ,
\ee
\be
\la{spiii}
{\J_1\ov w_1^3}+ {\J_2\ov w_2^3}+{\J_3\ov w_3^3}={1\ov
w_1^2w_2^2w_3^2}\int_{0}^{2\pi}\frac{d\sigma}{2\pi} \ \zeta_1\zeta_2\,
\ .
\ee
To find  the energy \eq{enn} in terms of the spins $J_i$
we need to express
 the frequencies $w_i$ and the Neumann
 integrals of motion parameters $b_a$
 in terms of $\J_i$  and use \eq{enna}.
 After finding a periodic solution of
 (\ref{sep}), this
  reduces to  the problem of computing the two independent
integrals on the r.h.s. of
eqs. (\ref{spii}) and (\ref{spiii}).

\subsection{Moduli space of the multi-spin string solutions}

We are thus interested in finding  periodic finite-energy
solitonic solutions of  O(6) sigma model defined
on a  2-cylinder  that carry three  global  charges
$J_i$. They can be  parametrised by  the three frequencies $w_i$
(or $J_i$)  as well  by the  two integrals of motion $b_a$.
The five parameters $(w_i,b_a)$ may be viewed as coordinates
on a moduli space of these solitons.

Because of the closed string periodicity condition
in $\s$,  general solutions will be classified
by  two    integer ``winding number''  parameters $n_a$
which  will  be  related to  $w_i$ and $b_a$ after solving
the periodicity condition \eq{pep}.
 In general, there will   be several different
solutions for given values of $J_1,J_2,J_3$,
i.e. the energy of the string
$E$ will be a function not only of
$J_1,J_2,J_3$ but also of  the values of $n_a$.

Depending on the values of these parameters
(i.e. location in the moduli space)
one may find
different geometric types  of the resulting
rotating string solutions.
In particular,  the string may be  ``folded''
(with topology of an interval)
 or  ``circular''
(with topology of a circle).
 A folded string may be  straight (as in the one- and
 two-spin examples
 considered in \ci{gkp} and \ci{ft4})
 or bent (at one or several points)
  as in the general three-spin case discussed
 below in Section 3.
 A ``circular'' string  may  have the form
 of a round circle as in the two-spin and three-spin solutions
 of \ci{ft2,ft3}  or a more general
 ``bent circle'' shape as in the three-spin solutions
 of Section 4 below.

Before turning to the $S^5$ rotation case, it is
useful to review  how these different string shapes
appear in the case of a closed string rotating in flat
$R^{1,5}$ Minkowski space.
 In orthogonal gauge, string coordinates are given by
solutions of free 2-d wave equation, i.e. by combinations  of
$e^{i n(\tau \pm  \s)}$, subject  to the standard
 constraints
$\dot X^2 + X'^2 = 0, \ \dot X X'=0$.
For a closed string rotating in the
two
orthogonal spatial planes 12 and 34 and  moving along
the 5-th spatial direction we find (cf. \eq{emb})
\be
X_0= \k \tau \ , \ \ \ \ \
 X_1 + i X_2 = \ x_1(\s) \ e^{ i w_1 \tau } \ ,
\ \ \ \   X_3 + i X_4 = \ x_2(\s) \ e^{ i w_2 \tau} \ ,
\ \ \ \  X_5 = p_5 \tau  \ , \ee
with
\be \la{taak}
w_1= n_1 \ , \ \  w_2 = n_2 \ , \ \ \
 x_1  = a_1   \sin (n_1 \s) \ ,  \ \ \ \ \ \
x_2  = a_2  \sin [n_2 (\s + \s_0) ]  \ .   \ee
Here $\s_0$ is an arbitrary integration constant, and $n_a$
are arbitrary integer numbers and
 the conformal gauge constraint implies that
 \be   \k^2 = p_5^2 +  n^2_1
a_1^2 + n^2_2 a_2^2 \ . \ee
 Then the  energy,  the two spins and the 5-th component of
 the linear momentum are
\be
\la{eess}
 E= {\k  \ov \a'} \ , \ \ \
\ \  J_1 = {n_1 a^2_1 \ov 2 \a'}  \ , \ \ \
\    J_2 =  {n_2 a^2_2 \ov 2 \a'}\ ,   \ \ \
P_5 ={p_5  \ov \a'} \ , \ee
i.e.
\be
 E= \sqrt { p^2_5  + { 2 \ov \a'} ( n_1 J_1 + n_2 J_2 ) } \ . \ee
To get  the two-spin states on the leading  Regge trajectory
(having minimal
energy for given  values of the {\it two} non-zero
 spins) one  is to choose
 $n_1=n_2=1$ with $\s_0= {\pi \ov 2}$.

The shape of the string depends on the values
of $\s_0$ and $n_1,n_2$.
If $\s_0$ is irrational then the string always has a ``circular''
(loop-like) shape. In general, the ``circular''  string
will not be lying in one plane, i.e. will have one
or several bends.
For rational values of $\s_0$ the string can be
either circular or folded,  depending  on
the values of $n_1,n_2$.

Consider as the first example the case of  $\s_0=0$.
If  $n_1=n_2$ the string is folded and straight, i.e. have no bends.
Indeed, then
$X_1+ i X_2$ is proportional to $X_3 + i X_4$
and thus one may put the string in  a single  2-plane
by  a global O(4) rotation.
If both $n_1$ and $n_2$ are either even or odd and different
then the string is folded and has several bends
(in the 13 and 24 planes). For example, if $n_2 = 3 n_1$
then the folded string  is wound $n_1$ times and
has two bends (in this case $x_2 = x_1 (4x_1^2 -3)$).

Next let us  consider the case of $\s_0 ={\pi\ov 2 n_2}$.
If  $n_1=n_2$ the string is an ellipsoid,
 becoming a round circle in the
special case of $a_1=a_2$  \ci{ft2}.
The string is also circular if $n_1$ is even and $n_2$ is odd.
If, however, $n_1$ is odd and $n_2$ is even
the string is folded. For example, if $n_2 = 2 n_1$
then the folded string  is wound $n_1$ times and
has a single bend at one  point
(in this case $ x_2 = 1 - 2x^2_1$).

The structure of the soliton string solutions
in   curved $S^5$ case is analogous.
Indeed, the equations of motion of the Neumann system are linearized
on the Jacobian of the hyperelliptic curve (\ref{ggg}).
The image of the string in the Jacobian whose real connected part
is identified with the Liouville torus can wind around two
non-trivial cycles with the winding numbers $n_1$ and $n_2$
respectively (see Appendix A).

\vskip 15pt
\noindent
\begin{minipage}{\textwidth}
\begin{center}
\includegraphics[width=0.30\textwidth]{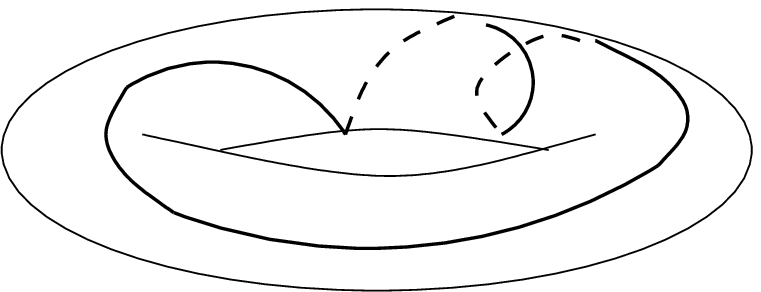}
\end{center}
\end{minipage}
\begin{center}
\parbox{5in}{
Fig.1: Image of a physical string at  fixed moment of  time
in the Jacobian (the Liouville torus). The string  winds
 around the  fundamental cycles
with  winding numbers $n_1$ and $n_2$.}
\end{center}
\vskip 20pt
The size and the shape of the Liouville torus
are governed by the moduli $(w_i,b_a)$. Specifying the winding numbers
$n_1,n_2$,  two of the five parameters $(w_i,b_a)$
are then  uniquely determined by
the  periodicity conditions. The actual shape of the physical string
 at  fixed moment of time  lying on two-sphere will depend
on the numbers $n_1,n_2$ and on the remaining moduli parameters
and may be of
(bent) folded type or of circular type.

Let us now  study
the folded and  circular string solutions in turn.

\setcounter{equation}{0}

\section{Folded  string solutions}

\subsection{Folded bent strings  with three  spins}

Our aim here will be to analyse folded string solutions
of \eq{sep}.
We shall  see that to have  three non-zero spins the
folded string must be bent at least at one point.

By definition,
a folded closed string configuration is such that
for all of the coordinates
$X_M (\s, \tau) = X_M(2\pi -\s, \tau)$, i.e.
$X_M'(0,\tau) =X_M'(\pi,\tau ) =0$
(we choose $\s= \pi$ as a middle point).
In  the case of  our rotating ansatz
\eq{emb} this  leads to
  $$x_i(\s) = x_i(2\pi -\s) \ , \ \ \ \ \ \ \
  x'_i (0) = x'_i ( \pi) =0, \ \ \ \ \ \ i=1,2,3\ . $$
If $x'_i$ vanishes for all $i$
 only at the two points, then the string has no
bends. Such straight folded string can carry only one
non-trivial component of the spin in flat space,
but in the case of rotation in  $ S^5$ it may carry two
non-zero spins \ci{ft4}.

To analyse
when the folded string can carry three non-zero spins
 let us use the angular $(\psi,\g)$
 parametrization of two-sphere
 formed by $x_i$ \eq{jio}.
Let us  consider a folded string stretched along
$\g$ and $\psi$,
\bea
\label{gapsi}
-\psi_0 \le \psi(\s)\le \psi_0\ ,\ \ \ \ \ \ \ \
{\pi\over 2} - \g_0 \le\g(\s)\le {\pi\over 2} + \g_0\ ,
\eea
and assume that
\be
\la{si}
\psi({\pi\over 2}) = \psi({3\pi\over 2})=0\ ,\ \ \ \ \ \ \ \
 \psi(0)  = -\psi(\pi) = \psi_0,\ \ \ \ \
 \psi'(0)=\psi'(\pi) =0 \ ,
\ee
\be
\la{asi}
 \g({\pi\over 2}) = \g({3\pi\over 2})={\pi\over 2} \ ,
\ \ \ \ \ \
\g(0) ={\pi\over 2} - \g_0\ , \ \ \ \ \ \ \
\g(\pi)={\pi\over 2} + \g_0\ , \ \ \ \ \ \ \
 \g'(0)=\g'(\pi) =0 \ .
\ee
This configuration is  a folded string without bends.
The case with two spins considered in \ci{ft4}
corresponds to $\g_0 =0$, so one  may  expect
that to  have a string with a  small non-zero
$J_3$ one needs to consider a
case with small $\g_0$.
However, it is possible to see that  this
no-bend case is not a genuine three-spin case --
there is a global SO(3) rotation that can be used to eliminate
$J_3$.
Indeed, for a consistent semiclassical state interpretation
 one has  to check  that only three components,
$J_1\equiv J_{12},\ J_2\equiv J_{34},\ J_3\equiv J_{56}$
of the SO(6) angular momentum  are
nonvanishing.
Assuming that the above string configuration
exists  when all the  frequencies $w_i$ are different,
 the angular momentum conservation of $J_{36}$
requires the vanishing of the following integral
(cf. \eq{jio})
\bea
\la{vanint}
\int_0^{2\pi} d\s\ x_2 (\s)\ x_3 (\s) =
\int_0^{2\pi} d\s\ \sin\g\ \cos \g \  \sin\psi \ .
\eea
However, it is easy  to see that for the
folded string configuration \eq{asi}
the integrand  here is positive for any value of $\s$.
Thus $J_{36}$ (and  also $J_{45}$) do not
  vanish.
  We conclude, therefore, that for the case of all $w_i$ different,
the above string solution does not exist.
Now consider the case  $w_2 = w_3$.
As soon as $w_2 = w_3$,
one can  rotate the folded string
to  place it entirely on the equator ($\g = {\pi \ov 2}$) of $S^2$
inside  $S^5$. Then
$x_3(\s) = 0$  and  $J_3=0$
for the transformed configuration.
The conclusion is therefore that the folded  straight string
configuration can only correspond to a two-spin case,
i.e. the periodicity conditions  should  imply
$w_3 = w_2$.

The reason  why in the no-bend case one is  able to
rotate the string  by a global SO(3) transformation
to set $\g = {\pi \ov 2}$  is that
the folded string should be stretching along a geodesic
(i.e.  some
oblique
big circle) of $S^2$ ($\psi,\g$)  part of  $S^5$.
This follows from the fact that both of the derivatives
$\g'$ and $\psi'$  vanish only at the same (two  ending)
points ($\s= 0, \pi$) of the folded string
(having a wiggle or a bend at some intermediate point
of the string would mean the vanishing of one of the derivatives
$\psi'$ or $\g'$ there).

We conclude that to find a non-trivial folded  string
solution with
three spins we need to admit a possibility of bends,
i.e. the  points on the string where one of the two
coordinates has zero $\s$ derivative while the other does not
(see also Appendix D).
For example, if $\g'$   changes its sign not only
at $\s=0,\pi$ but also at some $2n$ other points while
$\psi'$ does not
that would mean one has $n$ folds in $\g$ but only one fold in
$\psi$, implying the existence of $n$ bend points.
In what follows we will be considering the simplest and
most symmetric case
of a single bend point located in  the middle of the folded string,
i.e. with $\g'$ vanishing at
$\s_0 ={\pi\over 2}$ and $\s_0 ={3\pi\over 2}$;
 such  configuration is expected to have
  minimal energy for given values of the three  spins.

To have  a folded string with a single bend we will thus require
 $$ x_i'(0) = x_i'(\pi) =0 \ , \ \ \ \ \ \
 x_3'({\pi\over 2}) = x_3'({3\pi\over 2}) = 0\ . $$
 In terms of the coordinates $\zi$ and $\zii$
 in \eq{xizi} these conditions
  can be satisfied if
\be\la{hoh}
\zi({\pi\over 2}) = \zi({3\pi\over 2}) =w_2^2\ ,
 \ \ \ \ \ \ \zi'({\pi\over 2}) = \zi'({3\pi\over 2}) =0
 \ . \ee
  In view
of equations \eq{sep} the second  condition
is equivalent to
\be\la{heh}
 \zii({\pi\over 2}) = \zii({3\pi\over 2}) =b_2 \ . \ee
The conditions \eq{hoh}, \eq{heh} mean
that  there should exist  4 points  $\s_1, ..., \s_4$
located as
\be  0 < \s_1 < {\pi\over 2} \ , \ \ \
{\pi\over 2} < \s_2 < \pi\ , \ \ \ \
\pi < \s_3 = 2\pi - \s_2 < {3\pi\over 2}\ , \ \ \ \
{3\pi\over 2} < \s_4 = 2\pi - \s_1 < 2\pi \  \ee
at which
$\zii = w_3$, and $x_3=0$. Passing through these points $x_3$ changes its sign.
The exact positions of $\s_1, ..., \s_4$
are determined by the values
of the parameters $w_i$ and $b_a$ of the solution.
Note that the middle of the folded
string at $\s = {\pi\over 2}, {3\pi\over 2}$
is not located at
$x_1 = 1, \ x_2=x_3=0$.
Thus, at $\s = 0$ we have  $\zi=b_1$ and $\zii=b_2$. Increasing
$\s$, both $\zi$ and $\zii$ increase until at
$\s = \s_1$ the coordinate  $\zii$
reaches the point $w_3^2$.
Then  $\zii'$ changes its sign
and $\zii$ begins to decrease. At $\s = {\pi\over 2}$
the coordinate
$\zi$ becomes  $w_2^2$ and $\zii$ reaches
 the turning point $b_2$.
Next,   $\zi$ begins to decrease and $\zii$ begins
 to increase until
at $\s = \s_2$ the coordinate $\zii$
reaches the point $w_3^2$ where $\zii'$ changes its sign again.
After that both $\zi$ and $\zii$ decrease and at
$\s = \pi$ they reach the turning points
$\zi=b_1$ and $\zii=b_2$  (see Fig.2).

\vskip 15pt
\noindent
\begin{minipage}{\textwidth}
\begin{center}
\includegraphics[width=0.50\textwidth]{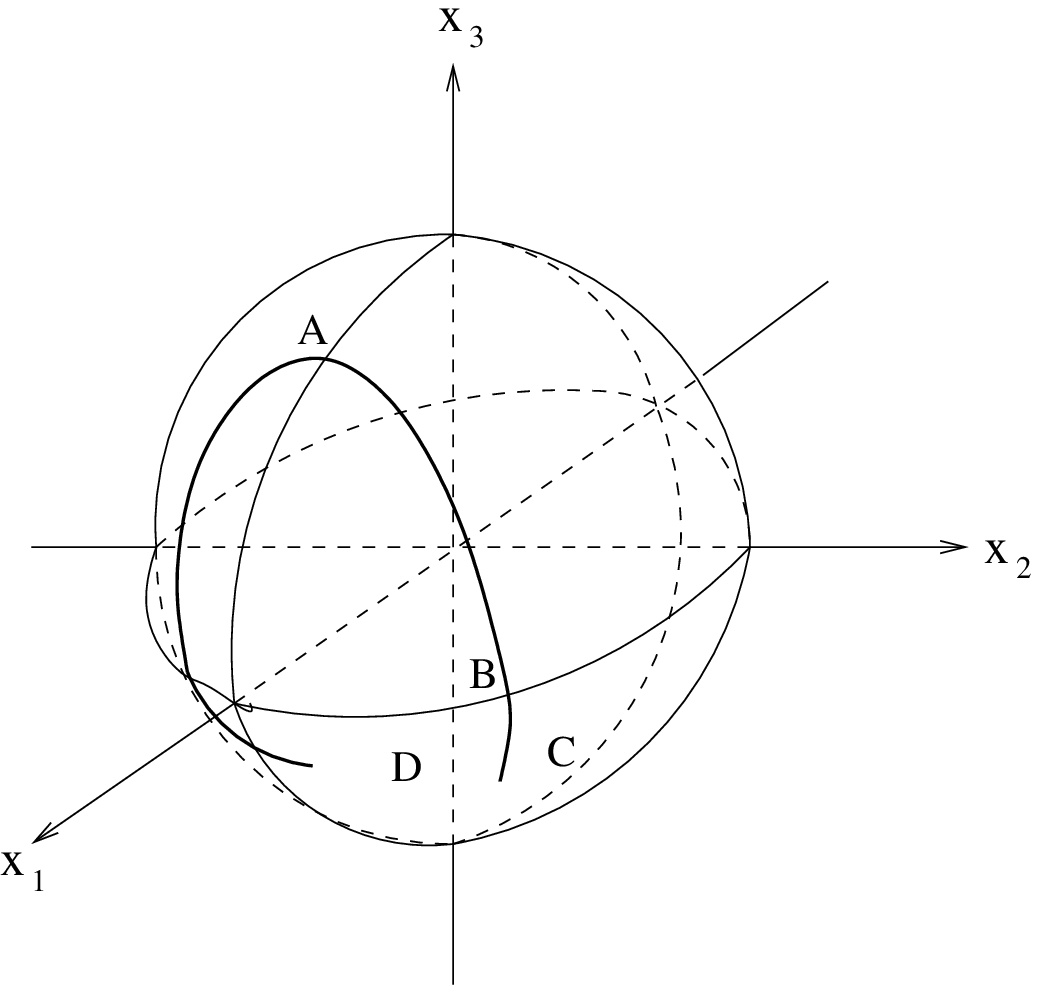}
\end{center}
\end{minipage}
\begin{center}
\parbox{5in}{
Fig.2:  Bent string.
The point $A$ corresponds to
$\sigma=\frac{\pi}{2},\frac{3\pi}{2}$,
the point $B$ - to  $\sigma=\s_1, 2\pi -\s_1$ and the points
$C$ and $D$ are  the turning points where $\sigma=0$
and $\s= \pi$. The turning points
$C$ and $D$ are symmetric w.r.t. to the plane 13.
When $b_2\to w_3^2$ the bend point $A$ tends to equator and
the string itself concentrates around $\gamma=\frac{\pi}{2}$,
i.e. we recover a straight string
 representing  the two-spin solution.}
\end{center}
\vskip 20pt

Thus, two   derivatives
$\z_1'$ and $\z_2'$ are both positive on the interval
$0 < \s <\s_1$,
and the equations of motion \eq{sep}  take the form
\bea
\label{zeta1i}
{d\zeta_1\over \sqrt{-P(\zi )}} =  2
{d\s\ov \zii-\zi}\ ,
\ \ \ \ \ \ \ \ \ \ \ \
{d\zeta_2\over \sqrt{-P(\zii )}} = 2
{d\s\ov \zii-\zi}\ .
\eea
The periodicity conditions \eq{pep}
follow from
(\ref{zeta1i}) and from the range \eq{r}, \eq{zibi}
of $\zi,\ \zii$ (these conditions do not depend on
$\s_1$)
\bea
\label{peri}
\int_{b_1}^{w_2^2}{d\zeta_1(\zii-\zi)\over \sqrt{-P(\zi )}} =
\pi\ ,
\ \ \ \ \ \ \ \
2\ \int_{b_2}^{w_3^2}{d\zeta_2(\zii-\zi)\over \sqrt{-P(\zii )}} =
\pi \ .
\eea
The presence of the coefficient 2 in the second equation reflects
the fact that we are considering the single-bend solution.
As a consequence of \eq{sep}
we have also the following relation
which is  valid for any point $\s$  from the interval
$0\le \s\le 2\pi$
\bea
\label{zizii}
|{d\zeta_1\ov d\zii}| = \sqrt{{P(\zi)\ov P(\zii)}}\ .
\eea
Then
\bea
\int_{c_1}^{w_2^2}{d\zeta_1\over \sqrt{-P(\zi )}}
&=&\int_{b_2}^{w_3^2}{d\zeta_2\over \sqrt{-P(\zii )}}
\ , \ \ \ \ \ \   c_1\equiv \zeta_1(\s_1)\ ,
\label{ppcc}
\\
\int_{b_1}^{c_1}{d\zeta_1\over \sqrt{-P(\zi )}}
&=&\int_{b_2}^{w_3^2}{d\zeta_2\over \sqrt{-P(\zii )}}
\nonumber
\label{ppccc}
\eea
 and,  therefore,
\bea
\la{zzizzii}
2\int_{b_2}^{w_3^2}{d\zeta_2\over \sqrt{-P(\zii )}}=
\int_{b_1}^{w_2^2}{d\zeta_1\over \sqrt{-P(\zi )}}
\ .
\eea
The conditions (\ref{peri}) and equations of
motion for $\zi, \zii$ \eq{zeta1i} also imply
\bea
\la{peri1}
2\int_{b_2}^{w_3^2}\frac{\zeta_2d\zeta_2}{\sqrt{-P(\zeta_2)}}
-\int_{b_1}^{w_2^2}\frac{\zeta_1 d\zeta_1}{\sqrt{-P(\zeta_1)}}=\pi
\eea
It will be convenient for the analysis in
next subsections to make
the following change of variables $\zeta_{1,2}\to \xi_{1,2}$
\bea
\la{changezi}
\zi = w_2^2-(w_2^2-b_1)\xi_1\ ,\ \ \ \ \ \ \ \
\zii = w_3^2-(w_3^2-b_2)\xi_2\ ,\
\eea
Then eqs. \eq{sep}  take the form
\bea
\la{dei}
\left(\frac{d\xi_1}{d\sigma}\right)^2
&=&4\frac{w_{21}^2(b_2-w_2^2)}{w^2_{32}}
\frac{\xi_1(1-\xi_1)(1-t_1\xi_1)(1-u_1\xi_1)(1-v_1\xi_1)}
{(1-u_1\xi_1-u_2\xi_2)^2}\ ,  \\
\la{deii}
\left(\frac{d\xi_2}{d\sigma}\right)^2
&=&
4\frac{w_{31}^2(w_3^2-b_1)}{w_{32}^2}
\frac{\xi_2(1-\xi_2)(1-t_2\xi_2)(1-u_2\xi_2)
(1-v_2\xi_2)}{(1-u_1\xi_1-u_2\xi_2)^2}\ .
\eea
Here we introduced the parameters
\bea
\la{tuvi}
t_1& =& {w_2^2-b_1\ov w_{21}^2} >0\ ,\ \ \ u_1=-{w_2^2-b_1\ov
w_{32}^2} <0\ ,\ \ \ v_1=-{w_2^2-b_1\ov
b_2-w_2^2} <0\ .  \\
\la{tuvii}
t_2 &=& {w_3^2-b_2\ov w_{31}^2} >0\ ,\ \ \ u_2={w_3^2-b_2\ov
w_{32}^2} >0\ ,\ \ \ ~~~v_2={w_3^2-b_2\ov
w_3^2-b_1} >0\ .
\eea

\subsection{System of equations for the leading
correction to the energy}

Let us now concentrate  on the case of solutions which have
all three  spins  very large \eq{asu}. The energy for such solutions
is expected to scale as in \eq{eee}.
Our aim is to derive a closed system of equations which
allows one to find the leading-order function  $f_1$ in \eq{eee},
thus determining
the  one-loop (first order in
$\l$) correction to conformal dimensions of the  corresponding
 dual SYM
operators with free-field theory dimension
$\Delta_0 = J_1+J_2+J_3$ in  SU(4) irreps with Dynkin labels
$[J_2-J_3,J_1-J_2,J_2+J_3]$.
The same system of equations should then be expected
to emerge from the thermodynamic limit of the algebraic Bethe
ansatz for the SO(6) spin chain derived from the SYM theory
\ci{mz1,mz2,bes2}.

The two-spin case \ci{ft4} and the
relation  \eq{enna}  for the energy
of the string suggest that one should look for solutions
which  have the following moduli parameters
\bea
\label{omi}
w_i^2 = \J^2_{\rm tot} + \w_i({\J_i\ov \J_{\rm tot}})
    + O ({ 1 \ov\J^2_{\rm tot}})
 \ ,\ \ \ \ \ b_a = \J^2_{tot} + \b_a({\J_i\ov \J_{\rm tot}})  +
  O ({ 1 \ov\J^2_{\rm tot}})  \ ,
\eea
where
\be \Jt = \sqrt{\l}\ \J_{\rm tot} = \sqrt{\l}\ (\J_1+\J_2+\J_3) \ , \ee
is the total spin of the string.
We shall assume that
in the large $\J_{\rm tot}$ limit
the ``corrections''   $\w_i$ and $ \b_a$
depend only on the ratios
$\J_i\ov \J_{\rm tot}$ = $J_i\ov \Jt$.
Note that the presence of a linear $O(\J_{\rm tot})$
term in $\w_i^2 $ and $b_a$ in \eq{omi}
can be ruled out by  ``parity'' ($w_i \to - w_i$, etc.)
 considerations.

Then the relation  \eq{enna} implies that
\be
\la{e2}
E^2 =\Jt^2 +\lambda (\w_1+\w_2+\w_3-\b_1-\b_2) +
O ({ 1 \ov\Jt})  \ ,
\ee
i.e.
\bea
\la{e22}
E =\Jt +{ \lambda \ov \Jt} f_1 ( {J_i \ov \Jt}) +
O ({ 1 \ov\Jt^3})  \ ,  \ \ \ \ \ \ \
f_1 =\ha (  \w_1+\w_2+\w_3-\b_1-\b_2) \ .
\eea
We want therefore to find  the expressions for
$\w_i$ and $\b_a$ and thus for $f_1$  in terms of
the current ratios  ${J_i \ov \Jt}$.

The five  parameters $\w_i,\b_i$ can be found by solving
a system of five equations that follows
in the limit $\J_{\rm tot}\to \infty$
from the periodicity conditions
(\ref{zzizzii}) and (\ref{peri1}),
equations (\ref{spi})-(\ref{spii}), and one of the
equations (\ref{spins}) expressing spins through the parameters.
This system can be written as follows in terms of the
``hyperelliptic'' integrals
$I_{10}, I_{20}, \I_{11}, \I_{21}, \I_{22}$
defined in Appendix B
\bea
\la{sysi}
&&2I_{20}=I_{10}\ , \\
\la{sysii}
&&2\I_{21} =\w_{32}I_{10}+ \I_{11} - \pi\ , \\
\la{sysiii}
&&\w_1 j_1 +\w_2 j_2 +\w_3 j_3 = 0\ , \ \ \ \ \ \ \ \ \
j_i\equiv {J_i\ov J_{\rm tot}}\ , \\
\la{sysiv}
&&\w_1 + \w_2 -\w_3 +{2\ov\pi}\left[{1\ov 2}\w_{32}^2 I_0
+ \w_{32} \I_{11} - {1\ov 2}(2\I_{22}-\I_{12})\right] =0\ ,\\
\la{sysv}
&&j_3 ={2t_2\ov \pi }\int_0^1 d\xi_1\, {d\s\ov
d\xi_1} \left( 1 -u_1\xi_1\right)\, \xi_2 (\xi_1 ) \ .
\eea
Here $\w_{ij}\equiv \w_{i}-\w_{j}$ and
\bea
\la{dsdzi}
{d\s\ov d\xi_1} = {\sqrt{\w_{32}}\left( 1- u_1\xi_1 -
u_2\xi_2\right)\ov
2\sqrt{\w_{21}(\b_2-\w_{2})}}{1\ov\sqrt{\xi_1(1 -
\xi_1)(1 - t_1\xi_1)(1 - u_1\xi_1)(1 - v_1\xi_1)}}  \ ,
\eea
which  follows from eq.(\ref{zeta1i}).
 We could use any of the
other two
 equations in (\ref{spins})
instead of eq.(\ref{sysv}).\foot{Let us also note that we cannot
use eq.(\ref{spii})
to determine the  leading order  correction
$f_1$  because at this order
eq.(\ref{spii}) is a consequence of (\ref{sysiii}) and
(\ref{sysiv}).}
All of the parameters in \eq{tuvi},\eq{tuvii}
and in the
above system of equations  depend
only on the ratios $j_i = {\J_i\ov \J_{\rm tot}} = {J_i\ov \Jt}$.

Since  the integral in
(\ref{sysv}) can not be computed analytically,
there are at least  two ways to proceed
for a generic three-spin case.
One may
 try  to compute this integral
 and therefore to solve the whole system \eq{sysi}--\eq{sysv}
 numerically (see Section 3.4).
 One may also try to
develop a  perturbation theory
around the two-spin solution of \ci{ft4}
assuming that the third spin component $J_3$ is small
as  compared to  the total spin $\Jt$.
Since the two-spin solution was expressed in terms of the
elliptic functions, one could expect that the same should be
true for a perturbative expansion  at  the vicinity
of the two-spin solution. This is indeed what happens as will be explained
in the next subsection.

\subsection{Two-spin solution and small $J_3$ expansion}

As was mentioned above,  the limit $b_2\to w_3^2$
should correspond to the two-spin solution. In this case the
folded string   stretches along the equator,
i.e. it is straight (without bends).
This limit corresponds to a  degeneration of
 the hyperelliptic curve \eq{ggg}
governing the string dynamics into an elliptic one.

To see explicitly how this happens
let us   change the  variables
$\zeta_{1,2}\to \xi_{1,2}$ as in (\ref{changezi}).
In the limit
$$\eps\equiv w_3^2-b_2\to 0$$
 one finds that
$t_2,u_2,v_2\to 0$ and $u_1\to v_1$ so that eqs. (\ref{dei}), (\ref{deii})
simplify  to
\bea
\la{deis}
\left(\frac{d\xi_1}{d\sigma}\right)^2
&=&4w_{21}^2\xi_1(1-\xi_1)(1-t_1\xi_1) \, , \\
\la{deiis}
\left(\frac{d\xi_2}{d\sigma}\right)^2
&=&
4w_{21}^2(1-u_1)\Big(1-\frac{t_1}{u_1}\Big)
\frac{\xi_2(1-\xi_2)}{(1-u_1\xi_1)^2} \, .
\eea
Now one can recognize in eq.(\ref{deis})
the differential equation for the elliptic $\mbox{sn}$
function, i.e. its  solution is
\bea
\la{ssn}
\xi_1=\mbox{sn}^2\left(\K(t_1)- \sqrt{w_{21}^2}\sigma,t_1 \right) \, .
\eea
Here $\K$ denotes the complete elliptic integral of
the first kind.\footnote{We recall the definitions of the elliptic integrals
in Appendix B.}
In writing eq. (\ref{ssn}) we have also used the fact that
$\xi_1'$ is negative on the interval $0< \sigma <\frac{\pi}{2}$
and that $\xi_1(0)=1$ due to the identity $\mbox{sn}(\K(t_1),t_1)=1$.
Eq. (\ref{ssn}) defines an elliptic curve with  modulus $t_1$.
Then  the second equation (\ref{deiis}) can be integrated to find
$\xi_2$. Note that
in the  two-spin case the variable
$\xi_2$ is an auxiliary one as  the physical
coordinates $x_{1},x_{2}$ ($x_3=0$, $\g= {\pi \ov 2}$)
 are parametrized in terms of $\xi_1$ only (cf. \eq{xizi},\eq{changezi}).
The simplicity of the two-spin case is thus related
to the fact that the equations for $\zi$ and $\zii$
decouple when $b_2= w^2_3$.

To determine
the parameters $(t_1,u_1)$ entering  eqs. (\ref{deis}) and (\ref{deiis})
from the periodicity conditions (\ref{zzizzii}) and (\ref{peri1})
let us study in more detail the relation (\ref{zizii}).
In general, the latter implies the following two equalities
\bea
\la{ziziii}
I_{1}(\zi ) = I_2(\zii ) + I_2(b_2 )\ ,&&\ \ \ b_1\le \zi\le c_1\ ,
\ \ \ b_2\le \zii\le w_3^2\ , \\
I_{1}(\zi ) = - I_2(\zii ) + I_2(b_2 )\ ,&&\ \ \ c_1\le \zi\le w_2^2\ ,
\ \ \ b_2\le \zii\le w_3^2\ ,
\eea
where we have introduced the two period integrals of the
Abelian differential of the first kind:
\bea
\label{abdifi}
I_{1}(\zi ) = \int_{\zi}^{w_2^2} {d z\ov\sqrt{-P(z )}} &,&\ \ \
\ I_{2}(\zii ) =\int_{\zii}^{w_3^2} {d z\ov\sqrt{-P(z )}}\ .
\eea
Making the change of integration variable $ z\to
w_2^2-(w_2^2-b_1)\xi_1\, z$ for $I_1$, and  $ z\to
w_3^2-(w_3^2-b_2)\xi_2\, z$ for $I_2$, and using (\ref{changezi}) we obtain
a more useful form of $I_1,I_2$ given in Appendix B.
In the limit $\eps=w^2_3 - b_2 \to 0$
both integrals can be easily computed
and one finds
\bea\la{Iziie}
I_{1}(\eps\to 0)=\frac{2\Pi
[u_1,\mbox{arcsin}\sqrt{\xi_1},t_1]}{\sqrt{w_{21}^2w_{32}^2 (b_2-w_2^2)}}
\ ,  \ \ \ \ \ \ \ \
I_2(\eps\to 0)={2\, \arcsin \sqrt{\xi_2}\ov
\sqrt{w_{31}^2w_{32}^2 (w_3^2-b_1)}} \, ,
\eea
where $\Pi$ denotes the incomplete elliptic integral of the third kind.
Now  (\ref{ziziii}) allows one to solve for $\xi_2$
in terms of $\xi_1$,
namely
\bea
\la{solxi2}
\xi_2(\xi_1)=\cos^2
\left[ \sqrt{(1-u_1)\Big(1-\frac{t_1}{u_1}\Big)}~
\Pi(u_1,\mbox{arcsin}\sqrt{\xi_1},t_1) \right]\ .
\eea
Since $\zeta_a\to b_a$ implies that $\xi_{a}\to 1$
we find the following
transcendental
equation relating the parameters $u_1$ and $t_1$:
\bea
\la{2spin1}
\Pi(u_1,t_1)&=&\pi\sqrt{\frac{u_1}
{(1-u_1)(u_1-t_1)}}\, .
\eea
Then eq. (\ref{solxi2}) can be written
 as
\bea
\xi_2(\xi_1)=\cos^2 \left[\pi
\frac{\Pi(u_1,\mbox{arcsin}
\sqrt{\xi_1},t_1)}{\Pi(u_1,t_1)}\right] \ ,
\eea
This relation is valid for all values of $\xi_2$ from
the interval $[0,1]$.
In particular, one recognizes that $\xi_2(1)=\xi_2(0)=1$ and
$\xi_2(c_1)=0$, where $c_1 \approx 0.56862$. This dependence of
$\xi_2$ on $\xi_1$ is a reflection of the $U$-shaped form of our string.

Let us now consider the periodicity condition (\ref{peri1}):
\bea
2I_{21}-I_{11}=\pi
\eea
where we have introduced the following period integrals
of the other Abelian differential of the first kind:
\bea
\la{abdifii}
I_{11}=\int_{b_1}^{w_2^2}\frac{z dz }{\sqrt{-P(z)}} \, \ ,\ \ \ \ \ \ \ \ \
I_{21}=\int_{b_2}^{w_3^2}\frac{z dz}{\sqrt{-P(z)}}\ .
\eea
The same change of variables as for (\ref{ziziii}) allows one to compute
these integrals in the limit $\eps\to 0$ with the result
\bea
\frac{\pi w_3^2}{\sqrt{w_{31}^2w_{32}^2(w_3^2-b_1)}}
-\frac{w_2^2}{u_1\sqrt{w_{21}^2w_{32}^4}}\Big[p\K(t_1)+
(u_1-p)\Pi(u_1,t_1) \Big]=\frac{\pi}{2} \, .
\eea
where $p=\frac{w_2^2-b_1}{w_2^2}>0$.
With the help of (\ref{2spin1}) the last equation reduces
to
\bea
\la{2spin2}
\K(t_1)=\frac{\pi}{2}\sqrt{w_{21}^2} \, .
\eea
Eqs. (\ref{2spin1}) and (\ref{2spin2}) completely determine the parameters
$t_1,u_1$ of the two-spin solution in terms of the frequencies $w_1,w_2$.
Note that the requirement $\xi_1(\frac{\pi}{2})= \xi_1(\frac{3\pi}{2})= 0$
produces the same eq. (\ref{2spin2}) for $t_1$.

In the limit $\eps\to 0$ we can compute the integral
\bea
\int_{0}^{\frac{\pi}{2}}(\zeta_1+\zeta_2)\, d\sigma
={1\ov 2}( 2I_{22}-I_{12}) \, ,
\eea
where
\bea
\la{periii}
I_{12}=\int_{b_1}^{w_2^2}\frac{z^2 dz }{\sqrt{-P(z)}} \, \ ,\ \ \ \ \ \ \ \
I_{22}=\int_{b_2}^{w_3^2}\frac{z^2 dz}{\sqrt{-P(z)}} \, .
\eea
As a result,
\bea
\int_{0}^{\frac{\pi}{2}}d\sigma\ (\zeta_1+\zeta_2)\,
=\frac{\pi}{2} (w_1^2+w_3^2)+\sqrt{w_{21}^2}\eE(t_1) \, ,
\eea
where $\eE$ is the complete elliptic integral of the second kind. Then
in the limit $\eps\to 0$ eqs. (\ref{spi}) and (\ref{spii})
reduce to the following equations
\bea
\la{2spin3}
&&{\J_1\ov w_1}+ {\J_2\ov w_2}=1\ , \\
\label{2spin4}
&& w_1 \J_1 +w_2 \J_2 -w_2^2=-
\frac{2}{\pi}\sqrt{w_{21}^2}\eE(t_1)\ ,
\eea
with  eq. (\ref{spiii}) being  their consequence.
Summarizing,  we have shown how  the three-spin hyperelliptic solution
degenerates to the two-spin elliptic one, the later being
completely determined
by the system (\ref{2spin1}), (\ref{2spin2}), (\ref{2spin3})
and (\ref{2spin4}).
The energy of the two-spin solution is determined from
\bea
\la{energy}
\E^2=
\kappa^2=w_1^2+w_{21}^2 t_1 \, .
\eea
In the two-spin case  with
$\J_1=\J_2\equiv \J $, $\J_3=0$
 we know  that \cite{ft4}
\bea
\label{sol}
t_1=0.826115+...\, , ~~~
w_1=2\J-\frac{0.272922}{\J} \, , ~~~~
w_2=2\J+\frac{0.272922}{\J} \ .
\eea
We can then  use (\ref{2spin1}) to determine  $w_3$. The result is
\bea
\la{w3}
u_1 = -0.777383\ ,\ \ \ \
w_{32}^2 = 2.32025\ ,\ \ \ \
w_{3}= 2\J + {1.70597\ov 2\J}\ .
\eea
Finally,
it  is now easy to compute the energy of the three-spin string up to
the term linear in $j_3\equiv {J_3\ov J_{\rm tot}}$. To this end we should
expand the system of equations (\ref{sysi})--(\ref{sysv}) in $\eps$ and use
the two-spin solution as the zero-order  approximation. Then
this
system reduces to a
linear system which can be readily solved.
We find that the parameters $\w_i,\b_1, u_1,t_1$ in \eq{tuvi}, \eq{omi}
have the following expansion in $j_3$:
\bea
\la{omilargeJ}
&&\eps\equiv w_3^2 - b_2 =\w_3 - \b_2= 6.12528\, j_3\ , \ \ \ \ \ \ \
j_3\equiv {J_3\ov J_{\rm tot}}\\
\nonumber
&&u_1= -0.777383 + 0.424835\,j_3\ , \ \ \ \
t_1= 0.826115 + 0.183849\, j_3\ , \ \ \ \ \\
\nonumber
&&\w_1 = -1.09169 - 3.86759\, j_3\ , \ \ \ \
\w_2 =1.09169 - 2.95629\, j_3\ ,\ \ \ \ \\
\nonumber
&&\w_3=3.41194 - 0.20351\, j_3\ , \ \ \ \
\b_1 = -0.712032 - 4.11054 j_3\ .
\eea
Using these  values of the parameters we find for the energy
\be
\la{eneroli}
\E^2 = \J_{\rm tot}^2 + \w_1 +\w_2 -\b_1 + \eps\ =
\J_{\rm tot}^2 + 0.712032 + 3.41194\, j_3\ , \ee
i.e. \be
\la{enerolii}
E^2 =J_{\rm tot}^2 + 0.712032\, \l + 3.41194\, {J_3\ov \Jt}\, \l\ .\ee
Thus
 the energy has the form \eq{eqq}
with positive coefficients \be \la{liii}
E = J_{\rm tot} +
0.356016 \frac{\l }{\Jt}\left( 1 + 4.79183{J_3\ov \Jt}\right)\ .
\ee
The coefficient $f^{(0)}_1 =0.356016 $ is the same as
in the two-spin solution
 \ci{ft4},\foot{This coefficient has a simple
origin: $f^{(0)}_1 = {1 \ov \pi^2} (2 t -1) [\K(t)]^2 $,
where $t=0.826115$ is the (unique)
 solution of  $\K(t) = 2 \eE (t)$.}
 while
$f^{(1)}_1= 4.79183$   is  the string-theory  prediction for
the term linear in $J_3$ in the one-loop
anomalous dimension of the dual CFT operator.

\subsection{Comments on    folded string solutions with general
values of $J_1,J_2,J_3$}

Let us now  study  more
general folded string solutions which can be far from
the two-spin configuration.
This can be done numerically as follows.
One starts with three values $w_1, w_2,w_3$ as input parameters
and solves the two periodicity conditions (\ref{zzizzii}), (\ref{peri1}),
 for the unknowns $b_1$ and $b_2$.
Then one determines the parameter $c_1=\zeta_1(\pi/4) $
by solving numerically eq.(\ref{ppccc}) (or, equivalently,
eq.(\ref{ppcc})~).
With these values $w_1,w_2,w_3,b_1,b_2,c_1$ one can then compute
\be
\int_0^{2\pi} {d\s \over 2\pi } (\zeta_1+\zeta_2) =
{2\over \pi } \int_{b_2}^{w_3^2}\frac{z^2 dz}{\sqrt{-P(z)}}-{1\over \pi }
\int_{b_1}^{w_2^2}\frac{z^2 dz }{\sqrt{-P(z)}}  \ ,\ \
\ee
\be
\int_0^{2\pi} {d\s \over 2\pi } \zeta_1\zeta_2 =
{1\over \pi} \int_{b_1}^{w_2^2}\frac{z\ z_2(z) \big( z_2(z)-z) dz }
{\sqrt{-P(z)}}
\ee
and find $J_1,J_2,J_3$ from eqs.~(\ref{spi})-(\ref{spiii}).
The parameters of  different solutions obtained in this way are
shown in Table 1.

\begin{table}[htbp]
\centering
\begin{tabular}{l|l|l|l|l|l|l|l|l|l}
$w_1^2$ & $w_2^2$ &
$w_3^2$ & $ b_1$ & $b_2$ &
$c_1$ &
$\J_1$ & $\J_2$ & $\J_3$ & $ \Delta \E^2$  \\
 \hline
23.52 & 25.80 & 28.29 & 23.87 & 27.81 & 24.70 & 2.20 & 2.32 & 0.48 & 1.06 \\
34.65 & 36.89 & 39.40 & 35.01 & 39.00 & 35.82 & 2.75 & 2.86 & 0.38 & 0.94 \\
47.63 & 49.88 & 52.40 &47.99 & 51.97 & 48.80 & 3.21 & 3.32 & 0.47 & 0.96 \\
1  & 4 & 9 & 1.14  & 4.64 & 2.31 & 0.26 & 0.59 & 1.33 & 3.45 \\
25 & 28 & 33 & 25.14 & 28.64 & 26.31 & 1.30 & 1.56 & 2.55 & 2.88  \\
49 & 52 & 57 & 49.14 & 52.64 & 50.31 & 1.82 & 2.13 & 3.35 & 2.83 \\
49  & 53 & 59 & 49.06 & 53.36 & 50.61 & 1.50 & 1.96 & 3.97 & 3.42 \\
49  & 55 & 64  & 49.01 & 55.09 & 51.33 & 1.16 & 1.59 & 4.96 & 4.47 \\
49  & 51 & 55 & 49.36 & 52.27 & 50.08 & 2.69 & 2.18 & 2.30 & 1.92 \\
\end{tabular}
\parbox{5in}{\caption{
Parameters for string configurations with different values of angular
momenta. \label{tab:energi}}} \end{table}

The first three entries are cases close to the two-spin case,
which are indeed consistent with
eq.(\ref{enerolii}) for the energy.
The input values of $w_1,w_2,w_3$ are obtained from
eqs.~(\ref{omilargeJ}) using $\J_3=0.5$, $\J_{\rm tot}=5$ for the first entry, and
$\J_3=0.4$, $\J_{\rm tot}=6$,  $\J_3=0.5$, $\J_{\rm tot}=7$,  for the second
and third entries.
According to eq. (\ref{enerolii}), the perturbation-theory values
for the correction to the energy
$$\Delta E^2 \equiv E^2-\Jt^2= \l \Delta \E^2  $$
found in the expansion in powers of $J_3\ov \Jt$
are, respectively, $\Delta \E^2\cong 1.04 $,
$\Delta \E^2\cong 0.96 $,  $\Delta \E^2\cong 0.94 $.
They  agree with the results in  Table 1.

The values of $J_1,J_2,J_3$ are also in good agreement
with direct  perturbation theory results.
Small differences  are expected,
in view of the higher order corrections in powers of
$J_3\ov \Jt$ and in view of the fact that $\Jt$ is not very large
(the results of Table 1 represent summation of all
terms in $1\ov \Jt$ expansion,
while the (\ref{enerolii}) contains only the leading correction term).
Other cases  in Table 1 are far from
the two-spin case, i.e. have  $J_3$ of the same order as (or larger than)
 $J_1, J_2$.

In general, a random choice of $w_1$, $w_2$, $w_3$
may not correspond to a folded string solution.
 For
large $J_i$, there are no folded string solutions when the $w_{ij}^2$
are not small compared to the $w_i^2$.

We have considered some cases with the same
values of $w_{31}^2, \ w_{21}^2$, but increasing
values of $w_i$. They exhibit the following interesting fact.
The differences $b_1-w_1^2$ and $b_2-w_2^2$,
$c_1-w_1^2$ are always the same. The difference in the  energy
$\Delta E^2$
approaches some asymptotic value
as $J_i $ increase.\footnote{For very large values of $J_i$
the difference
  $\Delta E^2\equiv E^2-\Jt^2$
should depend only on the ratios $J_i \ov \Jt$.}
For the particular  entries of Table 1, one observes that as the
differences $w_i^2-w_j^2$ increase, $b_1$ gets closer to
$w_1^2$ and $b_2$ gets closer to $w_2^2$.

In conclusion, there exist folded string solutions for diverse
values of $J_1,J_2, J_3$. In the case when $J_3$ is smaller
than $J_1,J_2$, the numerical calculation reproduces the
perturbation theory results of the previous subsection.

\setcounter{equation}{0}

\section{Three-spin string solutions of circular type}

\subsection{String solutions of circular type in ellipsoidal coordinates}

Let us start with recalling that our parameters  are assumed to
satisfy in general the conditions  \eq{r},\eq{zibi}, i.e.
\bea
\la{range}
w_1^2\le \zi\le w_2^2\le \zii\le w_3^2\ , \ \ \ \  \ \ \ \ \ \
b_1\le \zi\le b_2 \le \zii\ .
\eea
As was discussed in Section 3, to describe a folded string we
should consider  $b_1$ lying  in the same  range as $\zi$,
and
 $b_2$ lying  in the same  range as $\zii$,
 i.e.
\be \la{cond}
w_1^2\le b_1 \le w_2^2 \ ,\ \ \ \ \ \ \ \  \ \ \
 w_2^2\le b_2 \le w_3^2  \ .
 \ee
To find  a ``circular''  string solution
we should relax at least one of the conditions \eq{cond}, i.e.
to   assume that
either $b_1$ or $b_2$ do not belong to the corresponding intervals.
Thus there are two different cases to be considered
$$
I.~~~~w_1^2 < b_2 < w_2^2\ ,\ \ \ \ \ \ \ \ \ II.~~~~b_1 < w_1^2\ .
$$
Let us start with the case $I$.
We have then two  options for the value of  $b_1$:
$$
(i)~~~~w_1^2 < b_1 < w_2^2\ , \ \ \ \ {\rm  or} \ \
 \ \ \  \ \ (ii)~~~~b_1 < w_1^2\ .
$$
In what follows we shall  consider the case $(i)$, because
the case $(ii)$ appears to  similar to $II$.

When $w_1^2\le b_1 \le w_2^2$,  we have
$$
b_1\le \zi\le b_2\ , \ \ \ \ \ \ \ \ \ \ \ \  w_2^2\le \zii\le w_3^2\
$$
and there  are many different circular string configurations with these
values of parameters.
Let us first consider the simplest one corresponding to
$$b_1 = b_2\equiv b \ , \ \ \ \ \   { \rm i.e.} \ \ \ \ \ \ \
\zi=b \ .$$
Making the change of variable $\zii\to \xi_2 $,
$$
\zii = w_3^2 - w_{32}^2\xi_2\ ,
$$
we get from \eq{sep} the following equation for $\xi_2$
\bea
\la{eqxi2}
\left(\frac{d\xi_2}{d\sigma}\right)^2 = 4 w_{31}^2\xi_2 (1-\xi_2)
(1-
t\xi_2)
\ , ~~~~~t\equiv {w_{32}^2\ov w_{31}^2}.
\eea
Assuming the initial condition $\xi_2(0)=0$, the
 solution of this equation is
\bea
\la{sns}
\sqrt{\xi_2}=\mbox{sn}\Big(\sqrt{w_{31}^2}\sigma,~
t
\Big)\, .
\eea
According to this formula the function $\sqrt{\xi_2}$ oscillates
between -1 and 1 as $\sigma$ goes around the string.
Hence,  if we want our solution to describe a circular type
string  with a
winding number $n$, the  real period of
the function $\mbox{sn}$  should be equal to ${2 \pi \ov n}\sqrt{w_{31}^2}$, i.e.
\bea
\la{wc1}
2\pi\sqrt{w_{31}^2} = 4n\K(t
) \, .
\eea
We used the fact that the elliptic  function
$4\K$ is the real period of  the function
$\mbox{sn}$. Since eq.(\ref{wc1}) appears to be similar to
(\ref{2spin2})
it may not be  apparent why
we are dealing with strings of circular type
rather then with multifolded strings.
To clarify this point let us  write down the expressions for the
physical-space  coordinates $x_i$ in \eq{xizi}:
\bea
\nonumber
x_1=\sqrt{\frac{b-w_1^2}{w_{21}^2}}
\Big(\frac{w_{31}^2-w_{32}^2\xi_2}{w_{31}^2}\Big)^{1/2}\, , ~~~
x_2=\sqrt{\frac{w_2^2-b}{w_{21}^2}}(1-\xi_2)^{1/2} \, , ~~~
x_3=\sqrt{\frac{w_3^2-b}{w_{31}^2}}\xi_2^{1/2} \, .
\eea
When $\xi_2$ changes from $0$ to $1$ all $x_i$
remain non-negative. However, $x_2,x_3$ can acquire zero value, i.e.
can reach the boundary of the coordinate patch $x_i\geq 0$ and,
therefore,
 they   may   change  sign. Change of the sign means that
we go to another coordinate patch on $S^2$.
This is indeed the case as one can see by computing the derivatives
$x_{2,3}'$: One finds that $x_2$ changes its sign when $\xi_2$ crosses 1, while
the sign of $x_3$ changes when $\xi_2$ crosses zero. Thus, the shape of the
resulting string configuration appears to be of a {\it circular} type.

Conversely, for the folded string configurations of
Section 3, the coordinate $x_3$ is zero, while $x_1$ remains strictly
positive as $\xi_1$ changes from 0 to 1.
We see once again that the shape of the string
depends essentially on the relation between the parameters of the moduli
space (cf. Section 2.3) which
 explicitly occur in the expressions for
the coordinates $x_i$ in terms of  the variable $\xi_2$:
$x_i=x_i(\xi_2; w_i,b_a)$.

It is useful to note
that in the limit $w_3 \to w_2$ the periodicity condition
(\ref{wc1}) reduces to $w_{31}^2=n^2$ and thus we recover
the circular three-spin solution found in \cite{ft2}:
\bea
\sqrt{\xi_2}\ \ \to \ \ \
\mbox{sn}(n\sigma,0)\ = \  \sin n\sigma\, .
\eea

We want to emphasize that  neither the solution (\ref{sns})
nor the periodicity condition (\ref{wc1}) depend on
$b$. This parameter occurs only in the expressions of $x_i$
through $\xi_2$ and it tells us how many spins our solution has.
For generic $b$ we have a three-spin solution.
A two-spin solution
arises in the limit $b\to w_1^2$ (or $b\to w_2^2$).
In this limit, $x_1= 0$, i.e.
$\J_1=0$, while \bea x_2 =\sqrt{1-\xi_2} \, , ~~~~~x_3=
\sqrt{\xi_2} \, .\eea The spins $\J_2$ and $\J_3$ are then easy to
compute \bea
\label{spinJ2}
\J_2&=&w_2\int_0^{2\pi}\frac{d\sigma}{2\pi}(1-\xi_2)=
\frac{w_2}{t}\left[t -1+\frac{\eE(t)}{\K(t)}\right]\, ,\\
\label{spinJ3}
\J_3&=&w_3\int_0^{2\pi}\frac{d\sigma}{2\pi}\xi_2=
\frac{w_3}{t}\left[1-\frac{\eE(t)}{\K(t)} \right] \, ,
 \eea
where we made use of the explicit solution (\ref{sns}). These two
equations can be used to express $w_{2,3}$ through $\J_{2,3}$, while
the parameter $t={w_{32}^2\ov w_{31}^2}$ can be found from (\ref{wc1}). The energy is then
given by
\bea
\label{energy2spin}
\E^2=w_2^2+\frac{w_{32}^2}{t} \, .
\eea
We will not go here into the detailed analysis of the two-spin
circular type solution above, we will do it in Section 5,
where a comparison with the gauge theory results will be presented.

Let us consider now the limiting case when  $w_3 \to w_2$. We also assume
that $w_1^2 < b_1$ and therefore $b_1 < \zi < b_2$. First we perform 
the following change of variables
$$
\zi = b_2 - b_{21}\xi_1\ ,\ \  \ \ \ \
\zii = w_3^2 - w_{32}^2\xi_2\ , \ \ \ \ \ \ \
\ \ b_{21}\equiv b_{2}-b_{1}\ .
$$
Then taking the limit $w_3 \to w_2$ we get from \eq{sep}
the following equations for
$\xi_1,\xi_2$
\bea
\la{qxi}
&&\left(\frac{d\xi_1}{d\sigma}\right)^2 = 4h\xi_1(1-\xi_1)(1-t_1\xi_1)
\ ,\\
\la{qxii}
&&\left(\frac{d\xi_2}{d\sigma}\right)^2 =  4h(1-u_1)\Big(1-\frac{t_1}{u_1}\Big){\xi_2 (1-\xi_2)\ov
(1-u_1\xi_1)^2} \ ,
\eea
where we introduced the parameters
\bea
0<t_1=\frac{b_{21}}{b_2-w_1^2}<1\, , ~~~~u_1=-\frac{b_{21}}{w_2^2-b_2}<0 \, ,
~~~~h=b_2-w_{1}^2>0\ .
\eea
These equations have the same form as
(\ref{deis}) and (\ref{deiis}) governing the two-spin
solutions for the folded string but there is an important difference.
Now the variable $\xi_2$ enters the
parametrization of the physical coordinates $x_i$:
\bea
\nonumber
x_1^2&=&\frac{b_2-w_1^2}{w_{21}^2}(1-t_1\xi_1)\, , ~~~~~
x_2^2=\frac{w_2^2-b_2}{w_{21}^2}(1-u_1\xi_1)(1-\xi_2)\, , ~~\\
x_3^2&=&\frac{w_2^2-b_2}{w_{21}^2}(1-u_1\xi_1)\xi_2 \, .
\eea
It is easy to see that
the corresponding  shape of the string is again of
a circular type.

As was pointed out   in Section 2.3,  generic winding string  solitons are
parametrized by two integers $(n_1,n_2)$. To construct a
corresponding solution we assume that the variable
$\sqrt{\xi_1}$ oscillates between $-1$ and $1$ with  period
$\frac{2\pi}{n_1}\sqrt{h}$. As above, this gives
an equation that  determines the parameter $t_1$:
\bea
\la{wp1}
\K(t_1)=\frac{\pi}{2n_1}\sqrt{h} \, .
\eea
Then eq.(\ref{qxii}) can be  integrated to give
\bea
\la{es}
\sqrt{\xi_2}=\sin\Big[
\sqrt{h(1-u_1)\Big(1-\frac{t_1}{u_1}\Big)}
\int_0^{\sigma}\frac{d\sigma'}{1-u_1\xi_1(\sigma')}\Big] \, ,
\eea
where the initial condition $\xi_2(0)=0$ was assumed.

Suppose now that the variable $\sqrt{\xi_2}$ performs $n_2$ oscillations
between $-1$ and $1$ as $\sigma$ runs from 0 to $2\pi$.
For $\sigma=2\pi$
the integral on the r.h.s. of (\ref{es}) can be easily
evaluated by a  change of variables,
and we obtain the second elliptic function
equation to determine the parameter $u_1$:
\bea
\la{wp2}
\Pi(u_1,t_1)=\frac{\pi}{2}
\frac{n_2}{n_1}\sqrt{\frac{u_1}{(1-u_1)(u_1-t_1)}} \, .
\eea
Equations (\ref{wp1}) and (\ref{wp2}) generalize
(\ref{2spin2}) and (\ref{2spin1}) for arbitrary winding
numbers $n_1$ and $n_2$. At this point
it should be mentioned that imposition of the relation 
$w_3=w_2$ implies that the three spins $\J_i$ are not any more independent 
but rather satisfy an additional constraint. Indeed, our solution
is governed by four parameters $w_1,w_2,b_1,b_2$ which obey 
the five equations, three of them determine the spins $\J_i$ through 
these parameters and
the other two are the periodicity conditions (\ref{wp1}) and (\ref{wp2}). 
Therefore, one of the equations is in fact a constraint for $\J_i$.

The case of the ``round-circle'' string can be recovered by taking
first the limit $t_1\to 0$ and then  $u_1\to 0$. In this double
limit eqs. (\ref{wp1}) and (\ref{wp2}) reduce to
\bea
b_2-w_1^2=n_1^2\, , ~~~~~\ \ \ \ \ \ \ n_1=n_2 \, .
\eea

Let us consider now the case $II$ and show that it is quite similar to
the case  $I$.
This time we have the following range of parameters
\bea
b_1\leq w_1^2 < w_2^2 <b_2<w_3^2 \, .
\eea
and thus
\bea
w_1^2 \leq\zi\leq w_2^2 \, , \, \, ~~~ \ \ \ \ \
b_2\leq \zii\leq w_3^2 \, .
\eea
To demonstrate that we are dealing with string solutions of circular type
we consider the limit $b_2\to w_3^2$.
Performing a change of variables
\bea
\zi = w_1^2+w_{21}^2\xi_1\ ,\ \ \ \ \ \ \ \
\zii = w_3^2-(w_3^2-b_2)\xi_2\ .
\eea
and taking the limit $b_2\to w_3^2$ one obtains the same system of equations
(\ref{qxi}) and (\ref{qxii}) parametrised by
\bea
t_1=-\frac{w_{21}^2}{w_1^2-b_1}<0\, ,
~~~~u_1=\frac{w_{21}^2}{w_{31}^2}>0 \, ,
~~~~h=w_1^2-b_1>0.
\eea
The coordinates $x_i$ depend only on  the variable $\xi_1$
\bea
x_1=\sqrt{\xi_1}\, , ~~~~\ \ \ \ x_2=\sqrt{1-\xi_1}\, ,~~~~~ \ \ \ \
x_3=0\, .
\eea
Applying the same arguments  as above
we conclude that the form of the string is of a
circular type.


One  can easily develop a perturbation theory around
the circular type solutions by using the equations expressing
spins through $w_i$ and $b_a$. Assuming the
same ansatz for
these parameters as in \eq{omi}
one can obtain a system of equations which  should determine
the leading-order correction to the energy in  \eq{eee}, i.e.
a one-loop anomalous dimension for the corresponding SYM operators.
 Again, an equivalent  system of equations
should follow from the  Bethe ansatz approach on the field-theory side.


\subsection{String solutions
 of circular type in spherical coordinates}

As was already discussed, when all
the frequencies
 $w_i$ are different, the
ellipsoidal coordinates allow one to separate the variables and
formulate a closed system of equations that  determines the energy
$E$ as  a function of spins $J_i$.
However, when two of the
three frequencies
 coincide, the U(1) subgroup of O(3) in \eq{A} is restored and
it is useful
to adopt   the spherical coordinates. To make a contact with \cite{ft2}
in this and the following subsection we relabel $x_1\to x_3$, $x_3\to x_2$
and consider the case
when
\be \la{now}  w_1 = w_2 \   >  \ w_3  \ .   \ee

 The spherical coordinates
are global and thus  particularly appropriate
 for the study of
solutions with non-minimal energy that
represent  strings of
circular type.
 Here we shall first present the
equations for the   spherical coordinates and then  relate
them  to the above discussion of the circular type
strings in the ellipsoidal coordinates.

The equations of motion for the spherical coordinates $\psi$ and
$\g$ follow from the action (\ref{A}) and the metric (\ref{adm}),
and in the case $w_1=w_2$ take the form \bea\la{gameq} (\sin^2\g
\ \psi')' = 0\ ,\ \ \ \ \  \ \ \ \ \ \ \g'' + {1\ov 2}\sin 2\g \ (w_{13}^2
-\psi'^2)=0\ .
 \eea Then  \bea \la{eqpsi}
\psi'=\frac{c}{\sin^2\gamma} \ ,\ \ \ \ \ \ \ \
c=\const \ ,   \eea
 \bea
\gamma''-c^2\frac{\cos\gamma}{\sin^3\gamma}+\frac{1}{2}
w_{13}^2\sin 2\gamma=0\, \, ,  \label{gamit} \eea
and the  conformal gauge
constraint reduces to \bea\la{vir}
\gamma'^2&=&\kappa^2-\frac{c^2}{\sin^2\gamma}-
w_1^2\sin^2\gamma-w_3^2\cos^2\gamma \, . \eea
This can be rewritten as
\bea \la{tur}
x'^2= w_{13}^2
(x^2-a_-)(a_+- x^2 )\, , \ \ \ \ \ \ \ \ \
x\equiv x_3 = \cos \g \ ,
\eea where the constants
$a_{\pm}$ are \bea \la{apm}
a_{\pm}=\frac{1}{2w_{13}^2}\Big[w_{13}^2+w_1^2-\kappa^2 \pm
 \sqrt{(\kappa^2-w_3^2)^2-4c^2w_{13}^2} \ \Big] \, .
\eea Clearly, to have two different turning points $a_-$ and $a_+$
we have to impose a condition\footnote{This implies $w_3<w_1$,
\  $w_3<\kappa$.} $0<a_{\pm} \leq 1$. Note also that  $a_+=1$ for
$c=0$. The expressions of the integral of motion $c$ and the energy
of the system in terms of  the turning points $a_{\pm}$ are \bea \la{c}
c^2=w_{13}^2(1-a_+)(1-a_-) \, ,
~~~~~~\kappa^2=w_1^2+(1-a_--a_+)w_{13}^2\, . \eea
To establish a connection with the discussion
of the  circular type strings
 in terms of the  ellipsoidal coordinates, we introduce
a new variable
$$\xi=\frac{x^2-a_-}{a_+-a_-}\ .$$
 Then eq.
(\ref{tur}) takes  the same form as (\ref{qxi}) with parameters
\bea h=a_-w_{13}^2 \, ,\ \ \ \ \
 ~~~~t_1=-\frac{a_+-a_-}{a_-}<0\, . \eea
As to eq. (\ref{eqpsi}),  we rewrite it as \bea \la{psi}
\psi'^2=\frac{c^2}{(1-a_-)^2}\frac{1}{(1-u_1\xi)^2}\, \eea with
\bea 0<u_1=\frac{a_+-a_-}{1-a_-}<1 \, . \eea
Using
(\ref{c}),  one can establish a direct  correspondence
between  eq. (\ref{psi}) and eq.
(\ref{qxii}). The periodicity conditions are then given by
(\ref{wp1}), where $n_1=1$  and by (\ref{wp2}).
In this way  one
recovers the ellipsoidal coordinate
 description of the circular type
strings.


\subsection{Energy of  circular type strings}

In this subsection  we shall
 find numerical solutions of the periodicity conditions
for the circular strings and
 calculate their energy in the spacial case
of two equal frequencies \eq{now}.

Let us define the  two ``turning''
angles $\g_1,\ \g_2$ by
$$
a_-=\cos^2\g_1\ , \ \ \ \ \ \ \ \ \ \ \
a_+=\cos^2\g_2\ ,
$$
so that
$$
c=w_{13}\sin \g_1\sin \g_2\ , \ \ \ \ \ \ \ \ \ \
w_{13}=\sqrt{w_1^2-w_3^2} \ .
$$
The string extends from $\gamma=\g_1 $ (which is the value
closer to the equator point  $\g={\pi\over 2}$)
to $\gamma=\g_2$.

The string energy can be determined from \be
\la{enercirc}
E^2=\lambda \k^2=
\lambda
\big[ w_{13}^2(\sin^2\g_2-\cos^2\g_1)+w_1^2\big]\ .
\ee
Our aim  is to
express $E$  as a function of $J_3$ and $J_1+J_2$,
i.e. of
\be\la{ghg}
{\cal J}_3\equiv {J_3\over \sqrt{\lambda}}\ ,\ \ \ \ \ \
 \ \ \ \
{\LL}\equiv {\jL\over \sqrt{\lambda}}\ , \ \ \ \ \ \
\jL\equiv J_1 + J_2  \ .
\ee
{}The parameters $w_1, \ w_{13}$ are given by
\be \la{pop}
w_1= {\LL w_3\over w_3-\J_3 }\ ,\ \  \ \ \ \ \ \
\ \ w_{13}=w_3\sqrt{
{\LL^2\over (w_3-\J_3)^2}-1}\ ,
\ee
so we have three unknowns $w_3,\ \g_1,\ \g_2$ and three
equations:
\bea
{\pi w_{13}}=K_1(\g_1,\g_2) \ ,\ \ \ \
{J_3\over
\sqrt{\lambda }} = {w_3\over \pi w_{13}}\ K_2(\g_1,\g_2) \ ,\
\
\ \ {n\pi }=K_3(\g_1,\g_2)\ ,
\label{sistema}
\eea
where
$K_1(\g_1,\g_2) , \ K_2(\g_1,\g_2) , \ K_3(\g_1,\g_2)$ are the
integrals which in the previous subsection where computed in
terms
of the elliptic functions,
\be
K_1(\g_1,\g_2)=\int_{\g_1}^{\g_2} d\g {\sin\g \over
\sqrt{(\cos^2\g_2
-\cos^2\g )(\cos^2\g-\cos^2\g_1 )}} \ ,
\ee
\be
K_2(\g_1,\g_2)=\int_{\g_1}^{\g_2} d\g {\sin\g\  \cos^2\g \over
\sqrt{(\cos^2\g_2 -\cos^2\g )(\cos^2\g-\cos^2\g_1 )}} \ ,
\ee
\be
K_3(\g_1,\g_2)=\sin\g_1\sin\g_2 \int_{\g_1}^{\g_2} d\g
{(\sin\g)^{-1}\over \sqrt{(\cos^2\g_2 -\cos^2\g
)(\cos^2\g-\cos^2\g_1 )}} \ .
\ee
The system (\ref{sistema}) can
be reduced to a system of two equations and two unknowns
$\g_1,\
\g_2$ by noting that
$$
{\J_3\over w_3} = {K_2(\g_1,\g_2)\over K_1(\g_1,\g_2)}\ .\ \ \
 $$
Using \eq{pop}, the second and third equations of
(\ref{sistema}) become:
\be
{n\pi }=K_3(\g_1,\g_2)\ , \ \ \
\ \ \ K_2(\g_1,\g_2)= {\pi \J_3 } \sqrt{ {\LL^2\over \J^2_3[
{K_1(\g_1,\g_2)\over K_2(\g_1,\g_2)}-1]^2}-1}
\label{dose}
\ee
These can be
solved numerically for $\g_1$, $\g_2$. Note that for large
$\J_3$
the square root on the right hand side of the second equation
must be very small. This gives a hint for the values of the
parameters $\g_1,\ \g_2$ which solve these equations.
We shall always consider the case when
 $\J_3, \jL \gg 1 $.

\smallskip

The general features which result from
numerical analysis  are the  following.
 The above system has  solutions for all possible
$n=1,2,3,...$.
 For given $n$, there is a minimal  value of $J_3/\jL$.
A lower bound can be obtained for large $n$.
For $n\gg 1$, $K_3$ must be large,
which implies that the angles $\g_1, \g_2$ are close
to $\pi\over 2 $. In this region, one can approximate
the integrals, leading to the bound
$J_3/\jL > 1/4n^2$.

When $J_3/\jL$ is  small,
the solution has maximum eccentricity, with $\g_1 $ near
$\pi/2$
and $\g_2$ near 0.
 As $J_3/\jL$ increases, the angle $\g_1$
increases and $\g_2$ decreases, until some critical value of
$J_3/\jL$ where $\g_1=\g_2$. For higher values of $J_3/\jL$,
there is
no solution.

At the critical value, one has a circular string.
Indeed, since $\g (\sigma)$ in this case is constant, then $$\psi'=n=\const  $$
and
the solution reduces to the round-circle string solution
of  \cite{ft2,ft3}.
 The critical value of $J_3/\jL$ depends on $n$. To
compute
it, we take the limit $\g_2\to \g_1$ and large $J_3$ of the
equations in (\ref{dose}). This implies ${\jL\over J_3}\to {K_1\over K_2}-1.$
For $\g_1=\g_2\equiv \g_0$, one
has
$K_2/K_1\to \cos^2\g_0$, and $K_3\to \pi /(2\cos \g_0)$, so
the
equations become
\be \cos \g_0={1\over 2n}\ ,\ \ \ \ \ \ \ \
\tan^2\g_0={\jL\over J_3}\ . \label{condi} \ee
One also has
$w_{13}=n$. The second equation can also be written as
$\sin^2\g_0=\jL/(\jL+J_3) $, which coincides with the
 expression  for the angle
$\g_0$
of the
circular solutions obtained in \cite{ft3}. The first equation $\cos
\g_0={1\over 2n}$
implies that for given $n$, there is a single value of $\jL/J_3$
which gives circular solutions \be {\jL\over J_3}=4n^2-1 \ .
\label{lljj}
\ee
This is in contrast to \cite{ft3}, where there are
circular solutions for any $\jL/J_3$ at fixed $n$.
The origin of this extra condition
can be traced back to the condition
$K_1=\pi w_{13}$ coming from imposing periodicity of $\g
(\s)$.
In the circular solution of \ci{ft3} with $\g=\g_0=$const,
the
condition $\cos\g_0=1/(2n)$ does not appear, since
 constant $\g$ is already  periodic without  need to impose
extra
conditions.
The origin of this extra condition can also be understood by
considering perturbations around the $\g=$const  solution.
Expanding
$\psi = n \s + \td \psi(\s)$, \
$\g(\s)=\g_0+\td \g(\s )$,
and substituting into the general equation \eq{gamit}, we find
to first order
\be{ \td \g}''+(2n \cos\g_0)^2 \td \g =0 \  .
\label{flu}
\ee This has solution $\td \g(\s )=a \sin (m \s )$,\
 $m\equiv 2n
\cos\g_0$. Imposing periodicity of $\g_1 $, we find that $m$
must
be an integer. The solution with a single winding in $\g $ is
in
fact $m=1$, or $\cos\g_0=1/(2n)$, which is precisely the
condition
obtained above by taking the limit $\g_2\to\g_1 $ on the
general
solution.

Thus, for a given winding $n$ in  $\psi(\s )$, there exist
perturbations of the round-circle  solutions only
if  a circular
string is located at a special angle $\cos \g_0=1/(2n)$ (i.e.
for  special values of  $\jL/J_3$):
only
such  discrete set of  circular strings
 admit regular deformations.

\smallskip

Returning  to the general case, an  important feature
of the solution  is
that in the infinite $J_3, \jL$ limit with
 fixed $\jL/J_3$, the correction to the energy
$$
\Delta E^2 \equiv E^2-  (\jL+J_3)^2\ , \ \ \ \ \ \ \ \
\Delta \E^2 = \l^{-1} \Delta E^2 \
$$
is
 a function  (as in  \rf{e2})  of the ratio $\jL/J_3$
only \be E^2=
(\jL+J_3)^2+
\lambda \ff ({\jL\ov J_3})\ . \ee
Just as in the folded string case in Section 3.2,
this limit effectively singles out
the leading correction  to the energy
(\ref{enercirc}) (cf. \eq{eee})
\be E=\jL+J_3 + {\lambda
\over 2(\jL+J_3)}\ff({\jL\ov J_3})
+...
\label{expa}
\ee
A summary of  numerical results  for $n=1$ is given
 in Table 2.

\begin{table}[htbp]
\centering
\begin{tabular}{l|l|l|l|l|l}
${\J}$ & ${\J_3}$ &
$\J\ov \J_3$ & $ \g_1$ & $\g_2$ &
$\Delta \E^2 $  \\
 \hline
925  & 100& 9.25& 1.57061  & 0.17367 & 1.316\\
900 & 100 & 9  &  1.57055 & 0.17874 &  1.291  \\
850 & 100 & 8.5  & 1.57038 & 0.18987 & 1.242  \\
600 & 75 &  8 &  1.57010   & 0.20253 & 1.194 \\
400 & 50 &  8 &  1.57010   & 0.20253 & 1.194 \\
525 & 75 &  7 &  1.56879   & 0.23406 & 1.097 \\
350 & 50 &  7 &  1.56879   & 0.23406 & 1.097 \\
1500 & 250 & 6 & 1.56485 & 0.27827 & 1.002\\
500 & 100 & 5 & 1.56894  & 0.34622 & 0.910 \\
450 & 100 & 4.5 & 1.56740 & 0.39740 & 0.866 \\
2000 & 500 & 4 & 1.56427  & 0.47176 & 0.824 \\
800 & 200 & 4 & 1.56427   & 0.47176 & 0.824 \\
700 & 200 & 3.5 & 1.43237 & 0.59732 & 0.785 \\
1500 & 500 & 3 & 1.04763 & 1.04677 & 0.750 \\
\end{tabular}
\parbox{5in}{\caption{
Energies of string configurations with $n=1$,
 lying between  angles $\g_1 $ and $\g_2$. A few cases
with
equal $\jL/J_3$ are included to illustrate
 explicitly that for
these
large values of $\jL, J_3$ the angles and energies depend only
on
the ratio $\jL/J_3$. \label{tab:energ}}} \end{table}

\smallskip

{}For $\jL/J_3=3$, we see that the angles $\g_1 $ and $\g_2$ are
nearly equal. This means that $\g $ is approximately constant,
so
this is the case of the circular string mentioned above
lying at $\g=1.0476=0.3334\ {\pi}\approx  {\pi\over 3}$.

{}For the circular string solution with $\psi'=n$ const,
  the energy and the angle $\g$
are given by \cite{ft3}
\bea
 E^2_{\rm circ}=(\jL+J_3)^2+{\lambda n^2 \jL\over \jL+J_3} +...\ ,
\ \ \ \ \ \ \  \ \sin^2\g_0={\jL\over \jL+J_3}+...\ ,
\label{circo}
\eea
where dots stand for terms which vanish at large $J_3,\jL$.
In the case $\jL/J_3=3$, $n=1$, we find $\Delta \E^2_{\rm
circ}=0.75$,
and $\g_0=\pi/3$, which  is in full agreement with the case
$\jL/J_3=3$ of Table 2.

As a function of $\jL/J_3$, $\Delta E^2$ can be
roughly approximated by a straight line. A more accurate
fitting is by adding a $(\jL/J_3)^2$ correction (see fig. 3):
\be
\la{kk}
\Delta \E^2 =\ff({\jL\ov J_3})  \cong
c_0+c_1{\jL\over J_3}+c_2 { \jL^2\over J_3^2}\ ,\
\ee
\be \ c_0\cong 0.524\ ,\ \ \ \ \ \ \ \ \ c_1\cong 0.068\ , \ \
\ \ \ \ \ \ \ \ c_2=0.0002 \ . \ee
This formula has a  different  structure as compared to the
round-circle case,  eq.~(\ref{circo}).\footnote{It is
also possible to fit the data in terms
of an expansion in, say, $\jL/\Jt$ or $J_3/\Jt$,
$ \Jt= \jL + J_3$.
The formula (\ref{kk}) is given as a simple book-keeping
of the numerical data.}
The string configurations are in general very different. In
particular, as explained above,
for given $n$ there is a minimum value of $J_3/\jL$, so these
solutions do not include the circular
solution with $J_3=0$ surrounding the equator  $\g=\pi/2 $ as
a limit. The present solution with single ($n=1$)
winding exists only for $\jL/J_3$ lying
in the interval
between the minimal value at $\jL/J_3= 3$ and maximal value $\jL/J_3\cong 10$.
\vskip 10pt
\noindent
\begin{minipage}{\textwidth}
\begin{center}
\includegraphics[width=0.70\textwidth]{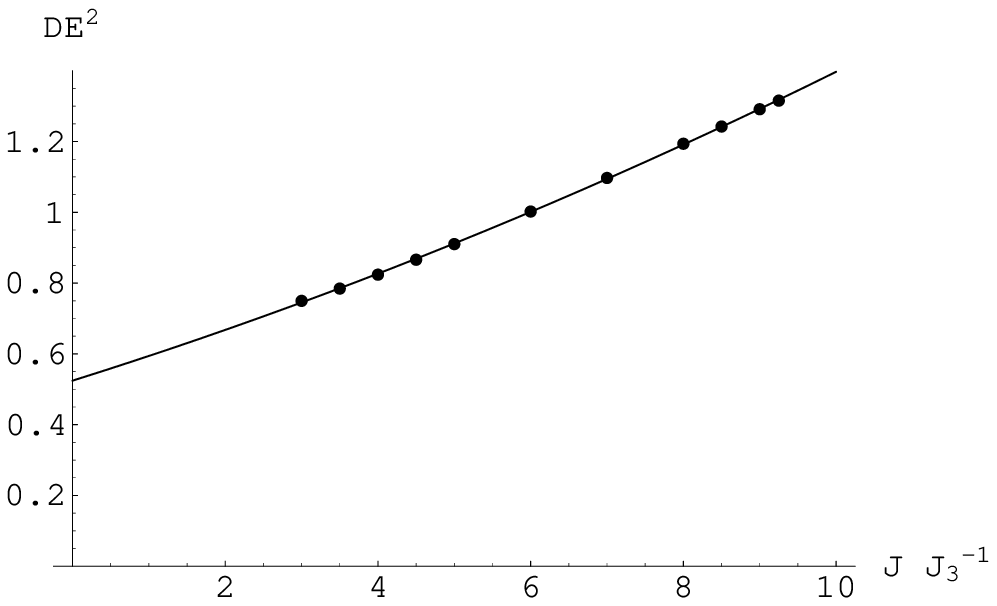}
\end{center}
\end{minipage}
\label{fig3}
\begin{center}
\parbox{7in}{
Fig.3:
Fitting of $\Delta E^2$ as a function of $\jL/J_3$ by
$\Delta \E^2 = c_0+c_1{\jL\over J_3}+c_2 { \jL^2\over J_3^2}$.}
\end{center}
\vskip 10pt

As a result,  we get the following expansion
for the energy of the corresponding string state
with
charges $\jL$ and $J_3$:
\be
E=\jL+J_3+ {\lambda\over 2(\jL+J_3) }
\big( c_0+c_1 {\jL\over J_3}+c_2 { \jL^2\over J_3^2} \big)
+O(\lambda^2)
\ee

\medskip

Let us  comment
on  solutions with higher winding numbers.\footnote{In general,
 one expects that for a string with fixed shape
(which requires fixed $\g_1,\g_2$) the energy grows as
$\Delta
\E^2\sim n^2...$. In the present case, different winding
numbers
give different angles $\g_1, \g_2$, so this simple $n^2$
dependence does not appear.}
For $n=2$ the critical value at which the string becomes
circular
is $\jL/J_3=4n^2-1=15$, and $\g_0=1.32 $ (this can be compared
with
the value $\g_0=\pi/3\cong 1.047$ appearing at $n=1$,
$\jL/J_3=3$).
For $n=3$, the critical value is $\jL/J_3=35 $, and $\g_0=1.40$,
etc.
{} The larger is $n$, the angles $\g_1 $, $\g_2$ get closer to the
value  $\g=\pi/2 $. In particular, for $n=2$ one
has solutions with:

\smallskip
\noindent (i) $\jL/J_3=10,\ \g_1=1.559 ,\ \g_2=0.791$,\  $\Delta
\E^2\cong 3.62$;

\smallskip
\noindent (ii) $\jL/J_3=12,\ \g_1=1.495 , \g_2=1.053$,\  $\Delta
\E^2\cong 3.69$.

\smallskip
The case (i) may be compared with the case $\jL/J_3=4$ in Table
1.
For a similar value of $\g_1$,  the
string is closer
to the equator, i.e. the higher is $n$,
the less is the eccentricity of the circular string.

\setcounter{equation}{0}

\section{Circular string with two angular momenta:\\
A  non-trivial test of AdS/CFT duality  }

Here we would like to complement the work  of Ref. \cite{ft4},
which considered
 the case of the {\it folded} string solution
 with two spins 
in $S^5$,
by discussing  the analogous  case of the circular type  two-spin
solution and comparing it with the corresponding SYM results 
{\mbox{of  \cite{mz2}.}

In Section 4 of Ref. \cite{mz2} a set of  SYM operators with
SU(4) Dynkin labels \mbox{$[J_2,J_1-J_2,J_2]$} was found, whose
one-loop anomalous  dimensions correspond to solutions of the
Bethe equations
 with all Bethe roots
lying on the imaginary axis. The one-loop anomalous dimension was
determined to be \be \label{zar} \Delta=J_1+J_2+ {\lambda \over
2\Jt} f(\eps ) \ ,\ \ \ \ \ \ \ \ \ \ \ \eps \equiv {1\over
2}-{J_2\over \Jt }={J_1-J_2\over 2(J_1+J_2)}\, , \ee where \be
\label{embo} f(\eps )=1+ 8\ \eps^2+ 24\ \eps^4 +96\ \eps^6+ 408\
\eps^8+... \ee Our aim here is to demonstrate that the
corresponding dual string configuration is described by a
circular type solution (\ref{eqxi2}), (\ref{sns}).

\medskip

To make a comparison with the SYM results straightforward it is
convenient to redefine the variables $w_i$ which obeyed $w_1\leq
w_2\leq w_3$ as follows: $w_1\to w_3$, $w_3\to w_1$. Then the
expressions (\ref{spinJ2}), (\ref{spinJ3}) and
(\ref{energy2spin}) transform into

\bea \label{spins'} 
\J_1=\frac{w_1}{t}\left[1-\frac{\eE(t)}{\K(t)} \right]
\, , ~~~~~~~
\J_2&=& \frac{w_2}{t}\left[t
-1+\frac{\eE(t)}{\K(t)}\right] \, \eea
and 
\bea \label{energy'} \E^2=w_2^2+\frac{w_{12}^2}{t} \, , \eea
where the parameter $t=\frac{w_{12}^2}{w_{13}^2}<1$. These equations are supplemented  by
the periodicity condition (\ref{wc1}) with $n=1$, which now reads as \bea
\label{period'} \K(t)=\frac{\pi}{2}\sqrt{\frac{w_{12}^2}{t}}. \eea

\medskip

Note that in spherical
coordinates $(\gamma,\psi)$ the equations of motion describing
our circular type string are $\gamma=\frac{\pi}{2}$ and $\psi'' +
\ha w_{12}^2 \sin 2\psi =0$. 
It is worthwhile to mention that after the obvious rescaling 
$\psi\to \frac{1}{2}\psi$ the last equation describes a motion of the 
plane pendulum in the gravitational field.
Integrating it once, we get
$\psi'^2 =w_{12}^2(1/t - \sin^2\psi )$, where $t$ appears as an
integration constant.  If $t>1$, then $1/t=\sin^2\psi_0$ and this
solution describes the {\it folded} string  extending from
$-\psi_0 $ to $\psi_0$.
For $t<1$, when there is no turning point where $\psi' =0$, the
solution will describe a circular string
 extending all the way around the equator $\g={\pi\over 2}$
with $\psi $ from $0$ to $2\pi $. In the limit  $t\to 0 $, this
solution will approach the circular type string
with $J_1=J_2$. Thus the parameter $t$ provides an interpolation
between the circular
and the folded string configurations. Clearly, the rotatory motion 
of the pendulum requires more total energy than the oscillatory phase 
and this explains why the energy of the circular string is bigger 
than that of the folded one. 

\medskip

We want to find the leading correction to the energy
(\ref{energy'}) in the limit of the large total spin $\Jt$ for a
generic circular type configuration with $J_1\neq J_2$ and to
determine thereby the one-loop anomalous dimension of the
corresponding SYM operators. To this end we assume the following
large $\Jt= \sqrt{\lambda}\J_{\rm tot}
 $ expansions
\bea w_1^2=\J^2_{\rm tot}+\w_1\, , ~~~~ w_2^2=\J^2_{\rm
tot}+\w_2\, , ~~~~ t\equiv t(\eps), ~~~\w_a=\w_a(\eps) \, ,\eea
where the variable $\eps$ is defined in (\ref{zar}). In the limit
of large $\Jt$ the energy (\ref{energy'}) is \bea
\E=\J_{\rm tot}+\frac{1}{2\J_{\rm tot}}\Big[\w_2+\frac{1}{t}(\w_1-\w_2)\Big]\, . \label{energy''} \eea
Expanding eqs. (\ref{spins'}) and (\ref{period'}) it is easy to
determine the $\w_{1,2}$ as the functions of the variable $t$:
\bea \nonumber
\w_1&=&\frac{4}{\pi^2}\K(t)\Big[(t-1)\K(t)+\eE(t)\Big] \, ,\\
\label{ws} \w_2&=&\frac{4}{\pi^2}\K(t)\Big[\eE(t)-\K(t)\Big] \,
, \eea while $t$ is determined as the function of $\eps$ from the
following equation: \bea
\eps=\frac{1}{t}-\frac{1}{2}-\frac{\eE(t)}{t\K(t)} \, . \eea The
last equation can be easily solved by the power series ansatz and
one obtains \bea \nonumber
t(\eps)&=&16 \eps - 128 \eps^2 + 736 \eps^3 - 3584 \eps^4 + 15808 \eps^5\\
&-&
    65024 \eps^6 + 253888 \eps^7 - 951296 \eps^8 + 3446272 \eps^9+...
\eea Inserting now this expansion in eqs. (\ref{ws}) we first
determine $\w_1$ and $\w_2$ and then the energy (\ref{energy''}).
The final result reads as 
\bea
E=\Jt+\frac{\lambda}{2\Jt}\Big[
\frac{4}{\pi^2}\K(t)\eE(t)\Big]
\eea
and one obtains the following well-defined perturbative expansion
of the energy
\be \label{gyi} E  = \Jt + {\lambda
\over 2\Jt } (1+ 8\ \eps^2+ 24\ \eps^4 +96\ \eps^6+ 408\
\eps^8+...\big)+... \ \  . \ee Quite remarkably, as in the folded
string case in  \cite{ft4},
 this exactly reproduces the full series (\ref{zar}), (\ref{embo})
for the anomalous dimension of the set of operators corresponding
to  imaginary Bethe roots in  \cite{mz2}.\footnote{The relative
position of the parameters $(w_i,b_a)$ describing the moduli
space of string solitons is therefore reflected on the field theory side
in  the way 
 the Bethe roots are distributed on  the complex plane.
Folded and circular type strings correspond to double contour
roots and imaginary roots respectively.}

For $\eps=0$ (i.e. for $J_1=J_2$), one recovers the expression
for the correction to the energy (anomalous dimension) ${\lambda
\over 2\Jt }$ corresponding to  the circular string solution
discussed in \cite{ft2}. The opposite, ``BPS-like'' limit
$\eps\to \frac{1}{2}$, i.e. $J_2\to 0$, is ill-defined.
This is in contrast to the folded string case, where 
$J_2\to 0$ leads to shrinking of the folded string to the BPS
particle (the corresponding dual operator with $\Delta=J_1$
transforms in the BPS irrep $[0,J_1,0]$). Obviously 
the circular string winding around the equator of $S^2$ can not be 
contracted to a point particle.

\medskip

We conclude this section by emphasizing that the Neumann dynamical
system and its string-like solutions encode the information about
the  full (all-loop) anomalous dimension of the corresponding
gauge theory operators. In fact,  our treatment above can be
extended in a straightforward manner to determine the subleading
terms in the energy which should correspond to the
 two- (and higher-) loop corrections to the anomalous dimensions of the
dual SYM operators. This might help to understand, from the
field-theoretic point of view, the integrability of the quantum
SYM theory beyond the one-loop level.

\setcounter{equation}{0}

\section{Strings rotating  in $AdS_5$}

In this section we briefly discuss
multi-spin  strings rotating in $AdS_5$ \ci{ft2},
emphasizing the similarity with the
case of rotation  in $S^5$.

We use the same rotation
 ansatz as in the $S^5$ case \eq{emb}  written in
terms of the embedding coordinates (\ref{rell}) as follows
\be
\la{emba}
Y_1+iY_2=y_1(\sigma)\ e^{iw_1\tau}\, , ~~~
Y_3+iY_4=y_2(\sigma)\ e^{iw_2\tau}\, , ~~~
Y_5+iY_0=y_3(\sigma)\ e^{iw_3\tau}\, .
\ee
The real radial functions
 $y_i$   are independent of time
and  should, as a consequence of $\eta_{MN}Y_M Y_N = -1$,
 lie on a two-dimensional hyperboloid:
\be\la{ett}  \eta_{ij} y_i y_j \equiv
- y_1^2 - y_2^2 + y_3^2 =1  \ . \ee
Note that because of the definition of $Y_5+iY_0$ in (\ref{rell}),
$t = w_3 \tau$, and, therefore, the 3-rd frequency
is now $w_3 = \k$.
Just as it was in the $S^5$ case, the three O(2,4)  spins
$$S_1=S_{12}\ ,\ \ \ \ \ \ \ \
 \ S_2=S_{34}\ ,\ \ \ \ \ \  \ S_3\equiv E=S_{56} \ $$
forming  a  Cartan subalgebra of  SO(2,4)  are
given by
\be
\la{spinsa}
S_i=\sqrt{\lambda}\ w_i
\int_0^{2\pi}\frac{d\sigma}{2\pi}\
 y_i^2(\sigma)\equiv\sqrt{\lambda}\ {\cal S}_i \, ,
\ee
and satisfy, because of \eq{ett},
the following relation
\be
\la{enerads}
{\E\over \k} - {{\cal S}_1\over w_1} - {{\cal S}_2\over w_2} =1 \ , \ \ \ \ \ \ \ \
\k= w_3 \ .
\ee
The effective 1-d mechanical system describing this
class of rotating  solutions  in $AdS_5$
has the following Lagrangian (obtained from \eq{SSL}
after an overall sign change, $L_{AdS} \to - \td L$)
\be\la{fo}   \td L= \frac{1}{2}\eta_{ij} ( y_i' y_j' - w^2_i y_i y_j)  +
\frac{1}{2}\td \Lambda ( \eta_{ij} y_i y_j -1)
 \ . \ee
Comparing eqs. \eq{fo},(\ref{spinsa}) and (\ref{enerads})
with the corresponding eqs. (\ref{L}),(\ref{spins})
and (\ref{spi}) for the $S^5$ case, we see
that the relation to the $S^5$ case is through
the analytic continuation\foot{The ``angular'' coordinates in
\eq{dam},\eq{adm}  are related by $\g \to i \rho$.}
\be \la{onn} x_1 \to i y_1\ ,  \ \ \ \ \ \ \ \
\ x_2\to i y_2   \ .  \ee
The ellipsoidal coordinates $\zeta_a$ defined as in \eq{ddd}
now provide the parametrization of the
two-sheeted hyperboloid $y_3^2-y_1^2-y_2^2=1$.
 Taking into account
the analytical continuation, we get the following relations between
$y_i$ and $\zeta_a$
\bea
\nonumber
y_1^2=\frac{(w_1^2-\zeta_1)(w_1^2-\zeta_2)}{w_{21}^2w_{13}^2}\,
,~~~~
y_2^2=\frac{(w_2^2-\zeta_1)(w_2^2-\zeta_2)}{w_{21}^2w_{32}^2}\, ,~~~~
y_3^2=\frac{(w_3^2-\zeta_1)(w_3^2-\zeta_2)}{w_{31}^2w_{32}^2}\ .
\eea
It is not difficult to check that the equations of motion for $\zeta_a$
and the Hamiltonian of the effective 1-d system
have the same form (\ref{sep}), (\ref{enna}) as
in the $S^5$ case. In terms of $\zeta_a$ the only difference between
the $S^5$ and $AdS_5$ cases is in the range of their allowed values.

Another essential feature of the string motion in $AdS_5$ space
is that in this case the Hamiltonian (\ref{enna})
coincides with the r.h.s. of the only non-trivial
 Virasoro constraint
$\eta_{MN}(\dot{Y}_M\dot{Y}_N + Y'_MY'_N)$, and, therefore, has to vanish
(recall that $\k = w_3$)
\be \kappa^2 + w_1^2+w_2^2-b_1-b_2 =0\ .\ee
This constraint, together with \eq{spinsa}
and periodicity conditions
  will  lead to the expression for the energy
as a function of the two (in general, unequal)  SO(2,4) spins,
$$ E= E(S_1,S_2)  \ . $$
The previously known
examples of the one-spin folded string solution
(with $w_1\not=0, w_2=0$)  \cite{vog,gkp}
 and the
circular string solution (with $w_1=w_2\not=0$) \cite{ft2} suggest  that
the frequencies $w_i$ should be chosen as
\bea
\la{w3w1w2a}
\k = w_3 \leq  w_1 \leq  w_2\ .
\eea
We shall  also assume for definiteness that
\bea
\la{z1z2a}
\zi < \zii\ ,\ \ \ \ \ \ \ \ \ \ \ \   \ b_1 < b_2\ .
\eea
Then one can  show that
\be
w_1^2\le \zi\le w_2^2 \le \zii\ , \ \ \ \ \ \ \ \
 \ \ \zi\le b_1\le \zii\le b_2 \ .
\ee
Just as it was in the $S^5$ case, a folded string solution exists if
$b_1$ and $b_2$ belong to the
same $w_i^2$ intervals as  $\zi$ and $\zii$, respectively, i.e. if
\bea
\la{b1w1b2w2a}
w_1^2 \le \zi \le b_1 \le w_2^2 \le  \zii \le b_2 \
.
\eea
A two-spin folded string solution exists only if
 the string is  bent.
The periodicity
conditions for a bent folded string are similar to the ones in the
$S^5$ case.
The relations between spins and the energy also
have the same form as the $S^5$ case
relations (\ref{spi}-\ref{spiii}),
 with
the replacement
$$\J_1\to -{\cal S}_1\ , \ \ \ \ \ \ \ \
\J_2\to -{\cal S}_2\ , \ \ \ \ \ \ \    \J_3\to \E    \ . $$
It would be interesting to analyze the
resulting system of equations in the limit of large spins $S_1,S_2$.
This limit seems to correspond to a long folded string
 with a  large  bend.

Two-spin string solutions of circular type exist only if
both $b_1$ and $b_2$  belong to the same interval as $\zii$, i.e.
\bea
\la{b1w1b2w2b}
w_1^2 \le \zi \le w_2^2 \le  b_1  \le  \zii \le b_2 \ .
\eea
Again, there are two simple cases
in which the solutions can be analysed in detail:
 $(i)$ $w_1 = w_2$,
and  $(ii)$ $b_1 = b_2$.
The simplest  round-circle  string solution
found  in \cite{ft2} corresponds
to the case $w_1 = w_2$ and $b_1 = b_2$.

\section{Concluding remarks }

In this paper we have developed a unified treatment of the
rotating string   solutions in \adss\ based on the integrability
of the Neumann dynamical system. We have shown that generic
multi-spin solutions are naturally associated to the
hyperelliptic genus 2 Riemann surface. The shape
of the closed string at fixed time may be of a
 folded (straight or
bent) string and of a circular   type. This depends
on the two winding numbers $n_1,n_2$ and on the relative
values  of the
parameters describing the solution  moduli space.

We have also studied perturbation theory around the
simplest two-spin solutions in the direction of the non-zero
third
spin component and derived a leading correction to the energy
in  this case. This enabled us to make an explicit prediction
for the 1-loop anomalous dimensions of the corresponding gauge-invariant
operators in ${\cal N}=4$ SYM theory (see (\ref{liii})).
One may  hope that
a simple picture of the constrained harmonic oscillator motion
linearizing on the Liouville torus may  be also
 uncovered in the equations
governing the algebraic Bethe ansatz
for gauge-invariant operators on
the SYM side (cf. \ci{mz2,bes2}).

It would be  interesting  also to see if and how the ``hyperellipticity''
of the general three-spin solutions is related to the more
complicated nature of the dual CFT operators. Indeed, in comparison
to  the ``elliptic'' two-spin solutions where the dual
operators
(tr$[ (\Phi_1 + i \Phi_2) ^{J_1}  (\Phi_3 + i \Phi_4) ^{J_2}]$+...)
are made out of
hypermultiplets (in the ${\cal N}=2$ language),
the operators dual to genuine three-spin
 hyperelliptic string solitons
will also mix (beyond the one-loop level)
with the operators from the field-strength multiplet.

As we have shown above, a rotating string in the $AdS_5$  space
is described by a  ``non-compact'' version of the Neumann
dynamical system which has a simple  interpretation
in terms of the harmonic oscillator
constrained to move  on the 2-d hyperboloid. It is desirable to
study
integrability of this system in more detail,  and, in particular,
to determine the energy
as a  function of spins $E=E(S_1,S_2)$
for ``long string'' configurations. A most natural
interpretation of the corresponding dual SYM operators will be
in terms of  non-local Wilson loops \ci{km} , and
one may  hope to shed some  light on their integrable
structures (see, e.g.,  \ci{korm} and refs. there).
One  specific  open problem is to study a folded
bent string solution with two equal spins
$S_1=S_2$ which should have lower  energy than the
circular solution found in \ci{ft2}.

It would be  very  important
 to try to go beyond semiclassics
and identify the string states and field-theoretic operators
for small values of the  spins and energy
(dimension). One possible direction  could be
to develop a string-bit model type
 approximation of continuous
string world-sheet (cf. \ci{ver}). Indeed, in the CFT
an elementary field contributes a ``quantum''
of dimension and spin to a composite operator.
Long composite operators are then viewed as
made of many quanta,
 and a  wave approximation corresponds to
considering  excitations of the continuous string world-sheet.
To understand the AdS/CFT correspondence
beyond semiclassical approximation
 one needs to find an analogue
of ``quantum" of energy and spin on the string side.

A related problem is to see
if and how the integrability of the
classical bosonic SO(2,4)$\times$SO(6) sigma model can be extended
to the  \adss\  Green-Schwarz  superstring sigma model
(for a recent work in this direction
 addressing  integrability of the classical
supercoset \ci{mt} sigma model see \ci{ben}).

\section*{Acknowledgments}
We are grateful to  N. Beisert, M. Staudacher
and S. Theisen for discussions of some related issues.
The work of  G.A. was supported in part
by the European Commission RTN programme HPRN-CT-2000-00131 and
by RFBI grant N02-01-00695.
The  work of S.F. and A.T.  was supported by the DOE grant
DE-FG02-91ER40690. The work of A.T.  was also supported in part
by the  PPARC SPG 00613 and  INTAS  99-1590 grants and the Royal
Society  Wolfson award. The work of J.R. is
supported in part by the European Community's Human potential
Programme under the contract HPRN-CT-2000-00131, and by MCYT FPA,
2001-3598 and CIRIT GC 2001SGR-00065.
 J.R. would like to thank Imperial College
for hospitality during the course of this work
and acknowledges also the support of PPARC SPG 00613 grant.

\setcounter{section}{0}
\setcounter{subsection}{0}

\setcounter{equation}{0}

\appendix{General solution
of the Neumann\\  system
in terms of $\theta $-functions}
The general solution of the $n=3$  Neumann system
can be formulated in terms of the theta functions
defined on the Jacobian of the hyperelliptic genus 2
Riemann surface \ci{mam}.
Introduce the theta-functions with characteristics $\eta$:
\bea
\theta[\eta](z)=\sum_{n\in
{Z}^2
} {\rm exp}\Big[
2\pi i z(n+\eta')+i\pi\tau(n+\eta')^2+2\pi i \eta''(n+\eta')\Big]\, ,
\eea
where $z=(z_1,z_2)$ and $n=(n_1,n_2)$.
Here $\tau$ is a period $2\times 2$
symmetric matrix with the positive definite imaginary part.
The characteristic $\eta$ is a $2\times 2$ matrix
$\eta=(\eta',\eta'')$ made of two columns $\eta'$
and $\eta''$.

The  two normalized Abelian differentials of the 1-st kind
(here $a,b=1,2$)
\be
\int_{A_a}\ww_b=\delta_{ab}\, , \ \ \ \ \ \
~~~~\int_{B_a}\ww_b=\tau_{ab}
\label{normal}
\ee
can be written in the form
\be
\ww_a=p_a\di_1+ e_a\di_2\ ,
\ee
where
\be
\di_1={d\zeta \over s}\ ,\ \ \ \ \ \ \ \
\ \ \di_2={\zeta d\zeta \over s}\ ,
\ee
and $s(\zeta)$ is determined by  (\ref{ggg}).
The differentials $\di_a, \ a=1,2$, satisfy
\be
A_{ab}=\int_{A_a} \di_b\ ,\ \ \ \ \ \ \ \ \
 \ B_{ab}=\int_{B_a} \di_b\
\ee
with $\tau_{ab}=(BA^{-1})_{ab}$.
The normalization condition (\ref{normal}) relates
the coefficients $e_a, \ p_a$
to $A_{ab}, B_{ab}$ (see also below).

In particular, when $\zeta\to \infty$,\
the leading coefficient of the differentials
$\ww_a$  are $e_a$ which  are the frequencies of oscillation
on the Jacobian. The solution in \cite{mam}  reads
($i=1,2,3$)
\bea
\la{thetasol}
x_i^2(\sigma)=\frac{\theta^2[\eta_{2i-1}](z_0+\frac{1}{2}ie\sigma
 )\theta^2[\eta_{2i-1}](0)}{\theta^2[0](z_0+\frac{1}{2}ie\sigma
 )\theta^2[0](0)} \, .
\eea
Here
\be
{\eta}_1=\left(\begin{array}{cc} \frac{1}{2} & 0\\
 0 & 0\end{array}\right)\, , ~~~
{\eta}_3=\left(\begin{array}{cc} 0 & \frac{1}{2}\\
\frac{1}{2} & 0\end{array}\right)\, ,~~~
{\eta}_5=\left(\begin{array}{cc} 0 & \frac{1}{2}\\
 0 & \frac{1}{2}\end{array}\right)\, \,
\ee
are the half-periods
and $z_0$ is a constant vector (the initial condition) which,
without loss of generality, we choose to be zero.
The specific coefficient $1/2$ in front of $e$ in eq. (\ref{thetasol})
is related to the fact that the canonical variable $\psi_a$ conjugate
to $b_a$ obeys the linear equation of motion
\bea
\psi_a'=\{H,\psi_a\}=\frac{1}{2}\, ,
\eea
where $H$ is given by eq. (\ref{enna}).

Finally,
the identity $\sum_{k=1}^3 x^2_k =1$ is due to the Frobenius formula
(here $g=2$)
\bea
\sum_{k=1}^{g+1}\frac{\theta^2[\eta_{2k-1}](z)\theta^2[\eta_{2k-1
}](0)}{\theta^2[0](z)\theta^2[0](0)}=1\ .
\eea
We do not know the constant vector
 $e_a$ explicitly, but we can impose the
 periodicity condition:
$x_i(\sigma+2\pi)=x_i(\sigma)$. In general, solutions (\ref{thetasol})
are not periodic functions of $\sigma$
and achieve periodicity we have to require
\bea
\label{constr1}
\pi e_a=\tau_{ab} m_b +n_a  \
\eea
where  $m_a$ and $n_a$ are vectors with  integer  components.
In addition, if we are interested in the real solutions, then
the motion occurs on the real connected component
of the Jacobian identified with the Liouville torus.
In this case $e_a$ is real and we obtain the condition
\bea
\label{constr2}
\pi e_a=n_a  \
\eea
Since  $e_1=-A_{21}/(\det A)$,
$e_2=A_{11}/(\det A)$, the latter condition reduces to
\bea
n_1A_{11}+n_2A_{21}=0\ , \ \ \ \ \ \ \
n_1A_{12}+n_2A_{22}=\pi \, .
\eea
Written in the integral form these conditions  are
\be
\la{fort}
n_1\int_{A_1} {d\zeta\over s}  + n_2 \int_{A_2} {d\zeta\over s}=0\  ,
\ee
and
\be
n_1\int_{A_1} {\zeta d\zeta\over s}  +n_2\int_{A_2}
{\zeta d\zeta\over s}  =\pi \ .
\label{fort1}
\ee
In this way we have demonstrated that the general periodic solitons
are characterized by two integers $(n_1,n_2)$.

Thus, when $\sigma$
goes from $0$ to $2\pi$ the image of the string in the Jacobian
winds around the real circles $A_1$ and $A_2$ with winding numbers
$n_1$ and $n_2$ respectively. The different periodicity conditions
discussed in the main text can be obtained by picking in
eqs. (\ref{fort}) and (\ref{fort1}) the concrete values for $(n_1,n_2)$
and specifying the cycles in terms of the branch points $(w_i,b_a)$.
It would be interesting to see how the various elliptic solutions discussed
in the paper can be directly obtained from (\ref{thetasol}) by
degenerating the period matrix $\tau$.

\setcounter{equation}{0}

\appendix{Basic     integrals}
Here we recall the definitions
of $\K(t)$, $\eE(t)$ and $\Pi(u,t)$ that are
the complete elliptic integrals of the first, the second
and the third kind respectively:
\bea
\K(t)&=&\int_0^1\frac{dx}{\sqrt{(1-x^2)(1-tx^2)}}=\frac{\pi}{2}
~_2F_1\Big(\frac{1}{2}, \frac{1}{2};1;t\Big) \, ,\\
\eE(t)&=&\int_0^1
\frac{\sqrt{1-tx^2}}{\sqrt{1-x^2}}dx=\frac{\pi}{2}
~_2F_1\Big(-\frac{1}{2}, \frac{1}{2};1;t\Big) \, ,\\
\Pi(u,t)&=&\int_0^1 \frac{dx}{(1-ux^2)\sqrt{(1-x^2)(1-tx^2)}} \, ,
\eea
where for $\K(t)$ and $\eE(t)$ we also provided their expressions in terms
of the Gauss hypergeometric function.

The incomplete elliptic integral of the third kind is given by
\bea
\Pi(u,\phi ,t)&=&\int_0^{\sin\phi}
\frac{dx}{(1-ux^2)\sqrt{(1-x^2)(1-tx^2)}} \, .
\eea

In Section 3.3  we used  the following integrals
\bea
\la{Izi}
I_{1}= {\sqrt{\xi_1}\ov
\sqrt{w_{21}^2w_{32}^2 (b_2-w_2^2)} }\int_0^1
{d z\ov\sqrt{z(1 - \xi_1 z)(1 - t_1\xi_1 z)(1 - u_1\xi_1 z)(1 -
v_1\xi_1 z)}}\ , \\
\la{Izii}
I_{2} = {\sqrt{\xi_2}\ov
\sqrt{w_{31}^2w_{32}^2 (w_3^2-b_1)}}\int_0^1
{d z\ov\sqrt{z(1 - \xi_2 z)(1 - t_2\xi_2 z)(1 - u_2\xi_2 z)(1 -
v_2\xi_2 z)}}\ ,
\eea
as well as
\bea
\nonumber
I_{10} &=&\int_{b_1}^{w_2^2}{dz\over \sqrt{-P(z)}} =
{1\ov
\sqrt{w_{21}^2w_{32}^2 (w_{32}^2-\eps)}}\int_0^1
{d z\ov\sqrt{z(1 - z)(1 - t_1 z)(1 - u_1 z)(1 -
v_1 z)}}\\
\nonumber
&\approx& {1\ov
\sqrt{w_{21}^2}\, w_{32}^2 }\int_0^1
{d z\ov\sqrt{z(1 - z)(1 - t_1 z)}(1 - u_1 z)} \\
&+&
{\eps\ov 2
\sqrt{w_{21}^2}w_{32}^4}\int_0^1
{d z\ov\sqrt{z(1 - z)(1 - t_1 z)}(1 - u_1 z)^2}
\ ,
\eea
\bea
\nonumber
I_{20}&=&\int_{b_2}^{w_3^2}{dz\over \sqrt{-P(z)}} =
{1\ov
\sqrt{w_{31}^2w_{32}^2 (w_{3}^2-b_1)}}\int_0^1
{d z\ov\sqrt{z(1 - z)(1 - t_2 z)(1 - u_2 z)(1 -
v_2 z)}}\\
\la{int20}
&\approx& {\pi\ov
\sqrt{w_{31}^2w_{32}^2 (w_{3}^2-b_1)}}
+{\eps\pi((w_{3}^2-b_1)(w_{31}^2+
w_{32}^2)+w_{31}^2w_{32}^2)\ov 4(w_{31}^2w_{32}^2
(w_{3}^2-b_1))^{{3\ov 2}}} \ ,
\eea
\bea
\nonumber
I_{11} &=&\int_{b_1}^{w_2^2}{z \ dz \over \sqrt{-P(z)}} =
{1\ov
\sqrt{w_{21}^2w_{32}^2 (w_{32}^2-\eps)}}\int_0^1
{ [w_2^2 - (w_2^2-b_1)z] \ d z \ov\sqrt{z(1 - z)(1 - t_1 z)(1 - u_1
z)(1 - v_1 z)}}\\
\la{int11}
&=& w_2^2 I_{10} - \I_{11}
\ .
\eea
Here $\eps = w_3^2 - b_2$ and
\bea
\nonumber
\I_{11} &=&
{w_2^2-b_1\ov
\sqrt{w_{21}^2w_{32}^2 (w_{32}^2-\eps)}}\int_0^1
{d z\,  z\ov\sqrt{z(1 - z)(1 - t_1 z)(1 - u_1
z)(1 - v_1 z)}}\\
\nonumber
&\approx& -{u_1\ov
\sqrt{w_{21}^2}}\int_0^1
{d z\, z\ov\sqrt{z(1 - z)(1 - t_1 z)}(1 - u_1 z)}\\
&-&
{\eps\, u_1\ov 2
\sqrt{w_{21}^2}w_{32}^2}\int_0^1
{d z\, z\ov\sqrt{z(1 - z)(1 - t_1 z)}(1 - u_1 z)^2}
\ .
\eea
Also,
\bea
\nonumber
I_{21} &=&\int_{b_2}^{w_3^2}{dz\, z\over \sqrt{-P(z)}} =
{1\ov
\sqrt{w_{31}^2w_{32}^2 (w_{3}^2-b_1)}}\int_0^1
{d z (w_3^2 - \eps z)\ov\sqrt{z(1 - z)(1 - t_2 z)(1 - u_2
z)(1 - v_2 z)}}\\
\la{int21}
&=& w_3^2 I_{20} - \I_{21}
\ ,
\eea
where
\bea
\I_{21} &=&
{\eps\ov
\sqrt{w_{31}^2w_{32}^2 (w_{3}^2-b_1)}}\int_0^1
{d z\,  z\ov\sqrt{z(1 - z)(1 - t_2 z)(1 - u_2
z)(1 - v_2 z)}}\\
&\approx& {\eps\pi\ov 2
\sqrt{w_{31}^2w_{32}^2 (w_{3}^2-b_1)}}
\ .
\eea
Similarly, we define
\bea
\nonumber
I_{12} &=&\int_{b_1}^{w_2^2}{dz\, z^2\over \sqrt{-P(z)}} =
{1\ov
\sqrt{w_{21}^2w_{32}^2 (w_{32}^2-\eps)}}\int_0^1
{d z (w_2^2 - (w_2^2-b_1)z)^2\ov\sqrt{z(1 - z)(1 - t_1 z)(1 - u_1
z)(1 - v_1 z)}}\\
\la{int12}
&=& w_2^4 I_{10} - 2w_2^2\I_{11} + \I_{12}
\ ,
\eea
where
\bea
\nonumber
\I_{12} &=&
{(w_2^2-b_1)^2\ov
\sqrt{w_{21}^2w_{32}^2 (w_{32}^2-\eps)}}\int_0^1
{d z\,  z^2\ov\sqrt{z(1 - z)(1 - t_1 z)(1 - u_1
z)(1 - v_1 z)}}\\
\nonumber
&\approx& {u_1^2\, w_{32}^2 \ov
\sqrt{w_{21}^2}}\int_0^1
{d z\, z^2\ov\sqrt{z(1 - z)(1 - t_1 z)}(1 - u_1 z)}
\\ &+&
{\eps\, u_1^2\ov 2
\sqrt{w_{21}^2}}\int_0^1
{d z\, z^2\ov\sqrt{z(1 - z)(1 - t_1 z)}(1 - u_1 z)^2}
\ ,
\eea
and
\bea
\nonumber
I_{22} &=&\int_{b_2}^{w_3^2}{dz\, z^2\over \sqrt{-P(z)}} =
{1\ov
\sqrt{w_{31}^2w_{32}^2 (w_{3}^2-b_1)}}\int_0^1
{d z (w_3^2 - \eps z)^2\ov\sqrt{z(1 - z)(1 - t_2 z)(1 - u_2
z)(1 - v_2 z)}}\\
\la{int22}
&=& w_3^4 I_{20} - 2 w_3^2\I_{21}+ \I_{22}
\ ,
\eea
\bea
\nonumber
\I_{22} &=&
{\eps^2\ov
\sqrt{w_{31}^2w_{32}^2 (w_{3}^2-b_1)}}\int_0^1
{d z\,  z^2\ov\sqrt{z(1 - z)(1 - t_2 z)(1 - u_2
z)(1 - v_2 z)}}\\
&\approx& {3\eps^2\pi\ov 8
\sqrt{w_{31}^2w_{32}^2 (w_{3}^2-b_1)}}\approx 0
\ .
\eea
Then the periodicity conditions and the integral $\int d\s (\zi +\zii)$
in Section 3.2 take the form
\bea
\la{eqi}
2I_{20}=I_{10}\ ,
\eea
\bea
\la{eqii}
2I_{21}-I_{11} = w_{32}^2I_{10}  - 2\I_{21} + \I_{11} = \pi\ ,
\ \ \ \ \  {\rm or} \ \ \ \  \ \ \
2\I_{21} =w_{32}^2I_{10} + \I_{11} - \pi\ ,
\eea
\bea
\int_{0}^{\frac{\pi}{2}} d \s \  (\zeta_1+\zeta_2)
={1\ov 2}I_{22}-\ha I_{12} = \pi w_3^2 -{1\ov
2}w_{32}^4I_0 - w_{32}^2 \I_{11} +
{1\ov 2}(2\I_{22}-\I_{12})\ .
\eea
We also have the following periodicity condition
\bea\nonumber
\pi &=& \int_{b_1}^{w_2^2}{d\zi\, [\zii(\zi) - \zi]\over
\sqrt{-P(\zi)}} = w_{32}^2 I_0 + \I_{11}\\
\nonumber
& -&
{\eps\ov
\sqrt{w_{21}^2w_{32}^2 (w_{32}^2-\eps)}}\int_0^1
{d \xi_1\, \xi_2(\xi_1) \ov\sqrt{\xi_1(1 - \xi_1)(1 -
t_1 \xi_1)(1 - u_1 \xi_1)(1 - v_1 \xi_1)}}\\
&\approx&
{1\ov \sqrt{w_{21}^2}}\int_0^1
{d z\ov\sqrt{z(1 - z)(1 - t_1 z)}} = {2\ov \sqrt{w_{21}^2}} \K(t_1)\ .
\eea

\setcounter{equation}{0}

\appendix{Vanishing of ``non-Cartan'' \\
components of angular momentum}

The  SO(6) momentum components $J_{MN}$ written in terms
of the 6 embedding coordinates $X_M$ in  \eq{relx}
are
\be
J_{MN}= \sqrt{\lambda } \int_0^{2\pi } {d\s\over 2\pi }
\ (X_M\partial_\tau X_N  -  X_N\partial_\tau X_M)\ .
\ee
They are  conserved in time by virtue of the SO(6) symmetry
of the Lagrangian  \eq{SL}.
The ``diagonal" components
 $J_1=J_{12},\ J_2= J_{34}$ and $J_3=J_{56}$)
are clearly non-zero for the rotating string ansatz \eq{emb}
(as long as all $w_i\neq 0$),
 being proportional to an integral \eq{spins} of a
positive definite quantity $x^2_i$.

The  question we would like to address
is  what are the  conditions for the
vanishing of  all other components of $J_{MN}$.
First, assume that   $w_1\neq w_2$. Then one has, in particular,
\be
J_{13}=\sqrt{\lambda }\left[w_1\sin (w_1\tau) \cos (w_2\tau)-
w_2\sin (w_2\tau) \cos (w_1\tau)\right]
\int  _0^{2\pi } {d\s\over 2\pi }    x_1(\s) x_2(\s ) \ .
\ee
Since $J_{13}$ must be  time-independent on shell,
 it follows that
$\int  _0^{2\pi } {d\s\over 2\pi }    x_1(\s) x_2(\s ) $ must vanish
on the solution of the string equations of motion.
Similarly,  $J_{14}$, $J_{23},\ J_{24}$ must
also vanish, since they are
proportional to the same integral over  $\s $.
One reaches analogous conclusions
in   the cases $w_1\neq w_3$ and $w_2\neq w_3$.

The solution of Section 3 where the
string is folded in both $\psi $ and $\g $ is possible only if
all  $w_i$ are  different, and so all extra $J_{MN}$ components
are necessarily zero there.

The only case that may in principle lead to  non-zero
values for the  ``non-Cartan'' components of
 $J_{MN}$
is when some two of the three frequencies happen to be
 equal.
If $w_1=w_2$, then  $J_{13}$ and $J_{24}$ are  automatically zero.
{}For the components $J_{23}$ and $J_{14}$ one finds
(using \eq{relx})
\be
J_{23}=-J_{14}= - \sqrt{\lambda}\ w_1 \int  _0^{2\pi } {d\s\over 2\pi }
x_1(\s) x_2(\s ) =
- \sqrt{\lambda}\  w_1\int  _0^{2\pi } {d\s\over 2\pi }    \sin^2\g
\cos\psi\sin\psi \ ,
\label{runo}
\ee
which may potentially give a non-zero result.
Similarly, if  $w_1=w_3$,
\be
J_{25}=-J_{16}=- \sqrt{\lambda}\ w_1  \int  _0^{2\pi } {d\s\over 2\pi }
x_1(\s) x_3(\s ) =-
\sqrt{\lambda}\ w_1 \int  _0^{2\pi } {d\s\over 2\pi }
   \sin\g\cos\g \cos\psi \ ,
\label{rdos}
\ee
and if  $w_2=w_3$
\be
J_{45}=-J_{36}=- \sqrt{\lambda}\ w_2 \int  _0^{2\pi } {d\s\over 2\pi }
x_2(\s) x_3(\s ) =- \sqrt{\lambda}\ w_2 \int  _0^{2\pi } {d\s\over 2\pi }
   \sin\g\cos\g \sin\psi \ .
\label{rtres}
\ee
Let us study  the values of these  components in some special  cases.
First, let us  consider a string configuration which is
folded in $\g $:
when  $\s $ goes from $0$ to $\pi $
let $\g $ vary  between  $\g_1 $ and $\g_2$ and
 $\psi $ between 0 to $\pi $, and
when $\s $ goes from $\pi $ to $2\pi $
let $\g$ vary from  $\g_2$ to $\g_1$  and  $\psi $
from $\pi $ to $2\pi $.
Then the  integral (\ref{runo})
vanishes because
the contribution of the
 integration region  $0< \s < \pi/2$ cancels
against the contribution of the integration region
   from $3\pi/2  < \s < 2\pi $, while
the contribution  of   $\pi/2 < \s < \pi $ cancels against
that of
 $\pi< \s < 3\pi /2 $. In the  same way,
  the integral (\ref{rtres})
also gives zero as the integration region
 $0< \s < \pi $ cancels with the region $\pi < \s < 2\pi $,
where $\sin\psi $ has the opposite sign.
The only non-obvious case is the vanishing of the
integral (\ref{rdos}). If there is
an extra symmetry such as,  e.g.,
$\g(\pi-\s)={\pi\over 2}-\g(\s)$
(as  in the cases where the
string extends
from the equator to both sides in a symmetrical way),
 then the
integral vanishes by similar
symmetry arguments.
If $\g(\pi-\s )$ is not directly related to $\g (\s )$ in a simple
way, this integral needs to be performed explicitly.
{}For the ``circular'' solution
  considered in Section 4, with $\psi'=c/\sin^2\g $,
$J_{25}$ and thus the  integral  (\ref{rdos}) must vanish because
there  $w_1\neq w_3$.
We have  checked independently that (\ref{rdos})
 is indeed zero in this case.

\setcounter{equation}{0}

\appendix{Straight folded string}

Here we describe the two-spin solution realized by the folded
string without bend points. As will follow from our consideration,
such solution is rigid in a sense that it does not allow a deformation
in the direction of the non-zero third spin component.

We assume the same range of parameters as in Section 3, i.e.
\bea
w_1^2<b_1\leq \zi\leq w_2^2<b_2\leq \zii\leq w_3^2 \,
\eea
and perform the same change of variables (\ref{changezi})
with the subsequent two-spin limit $b_2\to w_3^2$, so that
equations (\ref{sep}) will be reduced to the system
(\ref{deis}), (\ref{deiis}). However, this time
we assume the existence of the two turning points,
at $\sigma=0$ and at $\sigma=\pi$, and no bend points.
This implies that both derivatives ${\zeta'}_a$ are negative
on the interval $0\leq \sigma \leq \frac{\pi}{2}$ and
positive for $\frac{\pi}{2}\leq \sigma \leq \pi$.
The periodicity conditions describing this situation are
\bea
\la{D1}
I_1(\zi ) &=& I_2(\zii )\ , \\
I_{21}&-&I_{11}=\pi \ ,
\eea
where $I_{1,2}$ are defined in (\ref{abdifi}) while
$I_{11}$ and $I_{21}$ are given by (\ref{abdifii}).
Considerations similar to those in Section 3.3. allow one
to determine $\xi_2$ as  function of $\xi_1$:
\bea
\xi_2(\xi_1)=\sin^2 \left[\frac{\pi}{2}
\frac{\Pi(u_1,\arcsin
\sqrt{\xi_1},t_1)}{\Pi(u_1,t_1)}\right] \, .
\eea
This time $\xi_2$ is the monotonic function of $\xi_1$
on the interval [0,1]. In particular, $\xi_2(0)=0$ and
$\xi_2(1)=1$.

In addition,  the parameters $u_1,t_1$ should obey the following two
equations
 \bea
\la{E1}
\Pi(u_1,t_1)&=&\frac{\pi}{2}\sqrt{\frac{u_1}
{(1-u_1)(u_1-t_1)}}\, , \\
\la{E2}
\K(t_1)&=&\frac{\pi}{2}\sqrt{w_{21}^2} \, .
\eea
Comparison with (\ref{wp1}) and (\ref{wp2}) shows that the
image of our solution on the Liouville torus is characterized
by the winding numbers $n_1=n_2=1$.

The relations between $w_i$
and $\J_i$ retain the same form (\ref{2spin3}) and (\ref{2spin4})
as they do not depend on $\xi_2$.
Solution of the eqs. (\ref{2spin2}), which is the same as (\ref{E2}),
and (\ref{2spin3}), (\ref{2spin4}) for $w_1,w_2$ and $t_1$,
is given by eq. (\ref{sol}). By using this solution we can now
infer the value of $u_1$ from eq. (\ref{E1}). One finds
$u_1=-\infty \, $, which implies that $w_2=w_3$.

It turns out that the perturbation theory around $u_1=-\infty$
is ill-defined as this is essential singularity. Moreover,
the function $\xi_2(\xi_1)$ ceases to be regular in the limit
$u_1\to -\infty$ as $\xi_2(\xi_1)=1$ for
$0<\xi_1\leq 1$ and $\xi_2(\xi_1)=0$ for $\xi_1=0$.
Therefore, we conclude that our folded string is ``rigid''
in the sense that it
cannot be bent to acquire a small amount of the
 third spin component
$\J_3$.

It should be emphasized that the variable $u_1$ does not enter
either the relations between $w_i$ and $\J_i$
or the expression
(\ref{energy}) for the energy. This parameter and  the function
$\xi_2$ play only an  auxiliary r\^ole for the two-spin solutions.
However, they both arise in degeneration of a certain hyperelliptic
solution and,  therefore,  they point out  a direction in which an elliptic
two-spin solution can (not) be deformed.


\end{document}